\documentclass[prl, aps, nofootinbib, twocolumn, amssymb, superscriptaddress, 10pt]{revtex4-2}

\usepackage{amsmath}
\usepackage{amssymb}
\usepackage{amsthm}
\usepackage{braket}
\usepackage{amsfonts}
\usepackage{enumerate}
\usepackage{latexsym}
\usepackage{color}
\usepackage{xcolor}
\usepackage{setspace}
\usepackage{blindtext}
\usepackage{dsfont}
\usepackage{mathrsfs}
\usepackage[normalem]{ulem}
\allowdisplaybreaks
\usepackage{mathtools}
\usepackage{lipsum}

\usepackage{tabularx}
\newcolumntype{Y}{>{\centering\arraybackslash}X}

\usepackage{stackengine}
\usepackage{bbold}

\usepackage{bm}
\usepackage{graphicx}
\usepackage{wrapfig}

\usepackage{hyperref}
\hypersetup{
pdfnewwindow=true, colorlinks=true,
linkcolor=blue, anchorcolor=blue,
citecolor=blue, filecolor=blue,
menucolor=blue, urlcolor=blue}

\usepackage{pifont} 

\newcommand{\beginsupplement}{
        \setcounter{table}{0}
        \renewcommand{\thetable}{S\arabic{table}}
        \setcounter{figure}{0}
        \renewcommand{\thefigure}{S\arabic{figure}}
        \setcounter{equation}{0}
        \renewcommand{\theequation}{S\arabic{equation}}
        \setcounter{section}{0}
        \renewcommand{\thesection}{\Alph{section}}
        \setcounter{subsection}{0}
        \renewcommand{\thesubsection}{\arabic{subsection}}
}

\newcommand{\vk}{{\mathbf{k}}}
\newcommand{\vp}{{\mathbf{p}}}
\newcommand{\vq}{{\mathbf{q}}}

\begin{document}

\title{Metallic electron crystals in rhombohedral graphene as a bridge to twisted WSe\texorpdfstring{$_2$}{2}}
\author{Wentai Deng}
\thanks{These authors contributed equally to this work.}
\affiliation{Department of Physics \& Astronomy,  Extreme Quantum Materials Alliance, Smalley-Curl Institute,
Rice University, Houston, Texas 77005, USA}

\author{Chenyuan Li}
\thanks{These authors contributed equally to this work.}
\affiliation{Department of Physics \& Astronomy,  Extreme Quantum Materials Alliance, Smalley-Curl Institute,
Rice University, Houston, Texas 77005, USA}
\affiliation{Rice Academy of Fellows, Rice University, Houston, Texas 77005, USA}

\author{Naoto Nakatsuji}
\thanks{These authors contributed equally to this work.}
\affiliation{Department of Physics and Astronomy, Stony Brook University, Stony Brook, New York 11794, USA}

\author{Fang Xie}
\thanks{These authors contributed equally to this work.}
\affiliation{Department of Physics \& Astronomy,  Extreme Quantum Materials Alliance, Smalley-Curl Institute, Rice University, Houston, Texas 77005, USA}
\affiliation{Rice Academy of Fellows, Rice University, Houston, Texas 77005, USA}

\author{Lei Chen}
\affiliation{Department of Physics and Astronomy, Stony Brook University, Stony Brook, New York 11794, USA}

\author{Qimiao Si}
\thanks{\href{qmsi@rice.edu}{qmsi@rice.edu}}
\affiliation{Department of Physics \& Astronomy,  Extreme Quantum Materials Alliance, Smalley-Curl Institute, Rice University, Houston, Texas 77005, USA}

\author{Jennifer Cano}
\thanks{\href{jennifer.cano@stonybrook.edu}{jennifer.cano@stonybrook.edu}}
\affiliation{Department of Physics and Astronomy, Stony Brook University, Stony Brook, New York 11794, USA}
\affiliation{Center for Computational Quantum Physics, Flatiron Institute, New York, New York 10010, USA}

\begin{abstract}
    Superconducting phases have recently been discovered both in twisted bilayers of transition metal dichalcogenide WSe$_2$ (tWSe$_2$) and in rhombohedral multilayer graphene (RMG).
    While both have generated extensive interest, they appear to be driven by qualitatively different physics.
    Here, we propose a unified picture in which the superconductivity in RMG arises from a strongly correlated metallic state with a spatial length scale that is compatible with the moir\'e length scale of tWSe$_2$. 
    Central to this  mechanism is a Wigner crystal phase in RMG that 
    is metallic, driven by a ``W''-shaped dispersion in the flat part of the underlying graphene band structure.
    We demonstrate this effect through calculations of Wigner crystal states in pentalayer RMG, focusing on dilute electrons in the quarter-metal regime relevant to experiment.    
    In a region of the phase diagram we find the Wigner crystal breaks three-fold rotation symmetry, which guides our construction of a $t$-$V$ model on a triangular lattice as an effective description of the low-energy electronic physics.  This construction suggests a  surprisingly similar low-energy description of RMG and tWSe$_2$.
    Broader implications for superconductivity in dilute carrier systems are also discussed. 
\end{abstract}

\maketitle

\textit{\color{blue} Introduction}---
Strong correlation physics has been developing in increasingly 
versatile materials platforms. The discovery of correlated insulators 
and superconductors in twisted bilayer graphene \cite{cao_correlated_2018, cao_unconventional_2018} established
two-dimensional (2D) van-der-Waals multilayer materials as highly tunable systems for realizing many-body phenomena.
In the recent past, twisted bilayer transition metal dichalcogenides
(especially WSe$_2$) \cite{Xia2024Unconventional, Guo2024Superconductivity,Xu2025Signatures,Xia2026Bandwidth,  Guo2026} 
and rhombohedral multilayer graphene (RMG) \cite{Han2025Signatures,nguyen2026coexisting, Okounkova2026Two-dimensional, krötzsch2026surface, qin2026stripe, hua2026multiknob}
have emerged as hosts for superconducting states. 
Realized at low electron densities and accessible via electrostatic gating, these 2D material platforms offer great experimental tunability, and exhibit rich phase diagrams.
Additionally, the superconducting phases in these systems usually appear in the vicinity of other correlated phases, which is an indicator of their unconventional nature.
Yet, while extensive theoretical studies have ensued about the superconductivity
of both tWSe$_2$ \cite{li2025topological,Xie2024Superconductivity,Christos2024Approximate,Guerci2024Topological,Wu2023Pair,Qin2024KohnLuttinger,Chen2025,Tuo2025Theory,Kuhlenkamp_Arxiv_2025,schrade2024nematic} and 
RMG \cite{Cea2022Superconductivity, Jimeno2023Superconductivity, Yang2024topological,Chou2025Intravalley,Parra2025Band,Geier2026Chiral,Jahin2026Enhanced,Yoon2026Quarter,Qin2026Chiral,Dong2026superconductivity,Christos2026Finite, Wang2026Chiral},
a unified framework connecting the two remains lacking.

The superconductivity in RMG appears qualitatively distinct from that of tWSe$_2$ for two reasons.
First, in both the tetralayer (R4G), pentalayer (R5G) and hexalayer (R6G) graphene cases \cite{Han2025Signatures, nguyen2026coexisting, Okounkova2026Two-dimensional, krötzsch2026surface, qin2026stripe,hua2026multiknob},
superconductivity develops in a quarter metal normal state, making it likely that the superconducting state breaks time-reversal symmetry.
However, even though the normal state of tWSe$_2$ is time-reversal invariant, it is believed that the superconducting state is also chiral \cite{li2025topological,Xie2024Superconductivity,Christos2024Approximate,Guerci2024Topological,Wu2023Pair,Qin2024KohnLuttinger,Chen2025,Tuo2025Theory,Kuhlenkamp_Arxiv_2025}. 

Another distinction between the superconductivity in tWSe$_2$ and in RMG systems is the absence of a moir\'e potential in the latter.
It is this issue that we address. 
Phenomenologically, in the phase diagram of electron filling and vertical displacement field, superconductivity occurs adjacent to a correlated insulating region that likely is a Wigner crystal (WC).
Given that systems at low electron densities, including RMG,
tend to host competing many-body states with small differences in free energy \cite{Bernu2001Exchange, Candido2004Magnetic}, there is a high likelihood that vestiges of the Wigner crystal operate in the superconducting regime.
This would generate an intrinsic length scale set by the average inter-electron spacing, comparable to the moir\'e length scale of tWSe$_2$.
Indeed, experimentally, there have been a host of recent experiments providing evidence for the existence of such a length scale in RMG \cite{Han2026Evidence,zhou2026competing,nguyen2026coexisting}.
Equally important, very recent quantum oscillation measurements have suggested a strong reconstruction of the Fermi surface as the electron density is lowered towards the superconducting regime~\cite{Kalantre2026Fermiology, Zheng2026Chiral}, reminiscent of strongly correlated bulk crystals undergoing a correlation-driven localization transition \cite{hu2024quantumc,KirchnerRMP}.

\begin{figure*}[t!]
    \centering                    
    \includegraphics[width=1.0\linewidth]{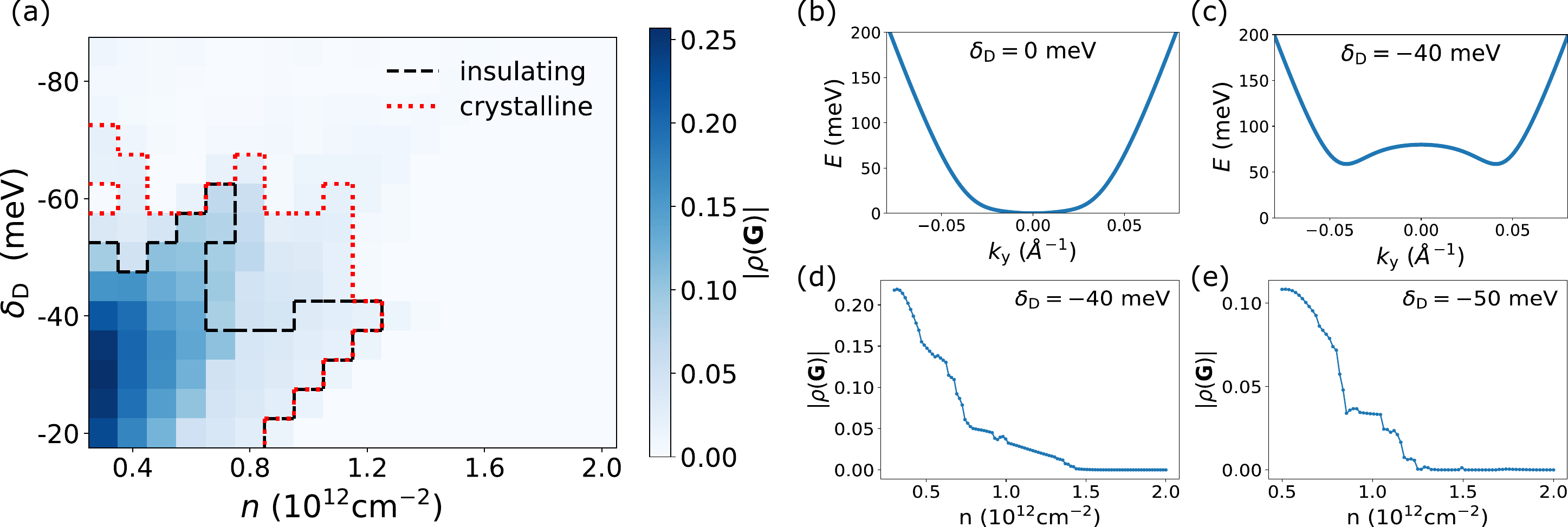}  
    \caption{(a) Phase diagram for Wigner crystallization as a function of the electron density $n$ and interlayer potential $\delta_{\mathrm{D}}$.
    $|\rho({\mathbf{G}})|$ 
    is the order parameter quantifying the broken continuous translation symmetry.
    The red and black dashed lines respectively mark the onset of 
    the crystalline phase $(|\rho({\mathbf{G}})|>0.02)$ and the metal-to-insulator transition.
    (b-c) The first conduction band of the noninteracting R5G band structure.
    The dispersion becomes ``W''-shaped for sufficiently large displacement field potential $\delta_D$, 
    leading to multiple Fermi surfaces.
    (d-e) Line cuts of $|\rho({\mathbf{G}})|$
    vs. $n$ for two fixed interlayer potentials.
    }
    \label{fig:rho_G}                              
\end{figure*}

In this work, we study the spontaneous emergence of such length scales in RMG. 
Central to our analysis is the interplay between the long-range Coulomb interactions in the dilute-carrier limit and the underlying crystalline potential at the atomic scale of graphene. 
The latter generates a ``W''-shaped dispersion at the flat band bottom of the conduction electron band [cf., Fig.~\ref{fig:rho_G}(c)], naturally enabling metallic Wigner crystal phases. To capture this behavior, we perform self-consistent Hartree-Fock calculations within a ten-band model of R5G, and obtain the correlated phase diagram as a function of vertical displacement field strength, parameterized by an interlayer potential $\delta_D$, and electron concentration ($n$).
Near the boundary between the WC and ordinary metallic states, we find metallic WC solutions with reconstructed Fermi surfaces. 
These metallic WC solutions, constructed on a triangular superlattice, can host a partially filled band that is either a trivial band or a Chern band. 
The resulting band structure thus has tunable topology on a triangular lattice, similar to continuum model band structures of tWSe$_2$ at different parameters.
This correspondence suggests that superconductivity in the two systems may be described within a common framework. We therefore construct a minimal $t$-$V$ model on the triangular lattice. The same model also captures the three-fold rotation ($C_{3z}$) symmetry breaking that we find in some of the metallic WC solutions.

The metallic WCs we present have one electron per unit cell;
in contrast, previous studies of commensurate Wigner crystals in RMG have focused on insulating states and their topology
\cite{dong2024anomalous,zhou2024fractional,zeng2024sublattice,dong2024stability,soejima2024anomalous,Crepel2025Efficient}.
While our work is distinct from proposals for ``self-doped" metallic  Wigner crystals with incommensuration between the electron filling and lattice spacing~\cite{Dong2026Crystals,Feng2026Selfdoped,Abouelkomsan2026Rhombohedral,Kim2026Shape},
we expect that a low-energy description with similar effective lattice models applies in those cases.
The $C_{3z}$-asymmetric phase we observe arises from a $C_{3z}$-symmetric unit cell, distinguishing it from the insulating $C_{3z}$-asymmetric lattices considered in Ref.~\cite{Ge2026Nematic};
this distinction is crucial for our advancement of the effective
low-energy description.

\textit{\color{blue} Model}---
We consider an interacting rhombohedral pentalayer graphene, described by the Hamiltonian
\begin{equation}\label{eqn:continuum-interacting-hamiltonian}
    H = H_{\text{R5G}} + \frac{1}{2A}\sum_{\mathbf{q}}V(\mathbf{q})\delta\rho_{\mathbf{q}}\delta\rho_{\mathbf{-q}}\,.
\end{equation}
Here, $H_{\text{R5G}}$ is a ten-band non-interacting continuum Hamiltonian describing a fully spin- and valley-polarized quarter metal,
which describes the long wavelength electronic structure of R5G around the $K$ point.
A vertical displacement field, parametrized by an interlayer potential $\delta_{\mathrm{D}}$,
strongly reshapes the low-energy band structure; it transforms 
 the conduction band bottom from a flat region 
 into a ``W''-shaped dispersion, as shown in Fig.~\ref{fig:rho_G}(b-c).
The electron-electron interaction we consider is a double-gate screened Coulomb potential. 
Further details are 
described in the {\it End Matter} and in Sec.~\ref{appsec:method} of the supplemental materials (SM) \cite{supplemental_material}.

We  perform self-consistent Hartree-Fock (HF) calculations,
employing a triangular lattice ansatz at commensurate filling with exactly one electron per unit cell.
For a given electron density $n$, the lattice constant is:
\begin{equation}
    a_{\rm WC} = \sqrt{{2}/({\sqrt{3}n})}\,.
    \label{eqn:latticeconstant}
\end{equation}
In the relevant parameter range, the lattice scale $a_{\rm WC}$ 
is on the order of $100\ \text{\AA}$, much greater than the value for the underlying graphene lattice.

\textit{\color{blue} Phase diagram}---
The crystalline order parameter  
$\rho(\mathbf{G})$
quantitatively characterizes the breaking of the translation symmetry.
The numerical criterion for crystallization
is taken to be $|\rho(\mathbf{G})|>0.02$, which is about 10\% of the maximal value in the phase diagram.
We find several crystalline states, as shown in Fig.~\ref{fig:rho_G}(a).
The crystallization is the strongest at low densities and is reduced upon increasing $n$. 
The red dotted line bounds the crystalline region. 
The density dependence of the crystalline order is further shown as line cuts at two representative values of $\delta_D$ in Figs.~\ref{fig:rho_G}(d-e). 
Within the phases with non-vanishing $|\rho(\mathbf{G})|$, both insulating and metallic states exist. The two solutions are separated by the black dashed line in Fig.~\ref{fig:rho_G}(a).

In the following, we will illustrate four representative solutions.
They correspond to distinct regions of the phase diagram within the crystalline bound.
For convenience, we will use four symbols -- filled and open triangles along with star and cross -- to mark the four parameter points on the phase diagram. They are specified in 
Fig.~\ref{fig:M5N12_parameter_points}, which also contains a phase boundary 
between regions with distinct band topology.

\begin{figure}[t]
    \centering                    
    \includegraphics[width=\linewidth]{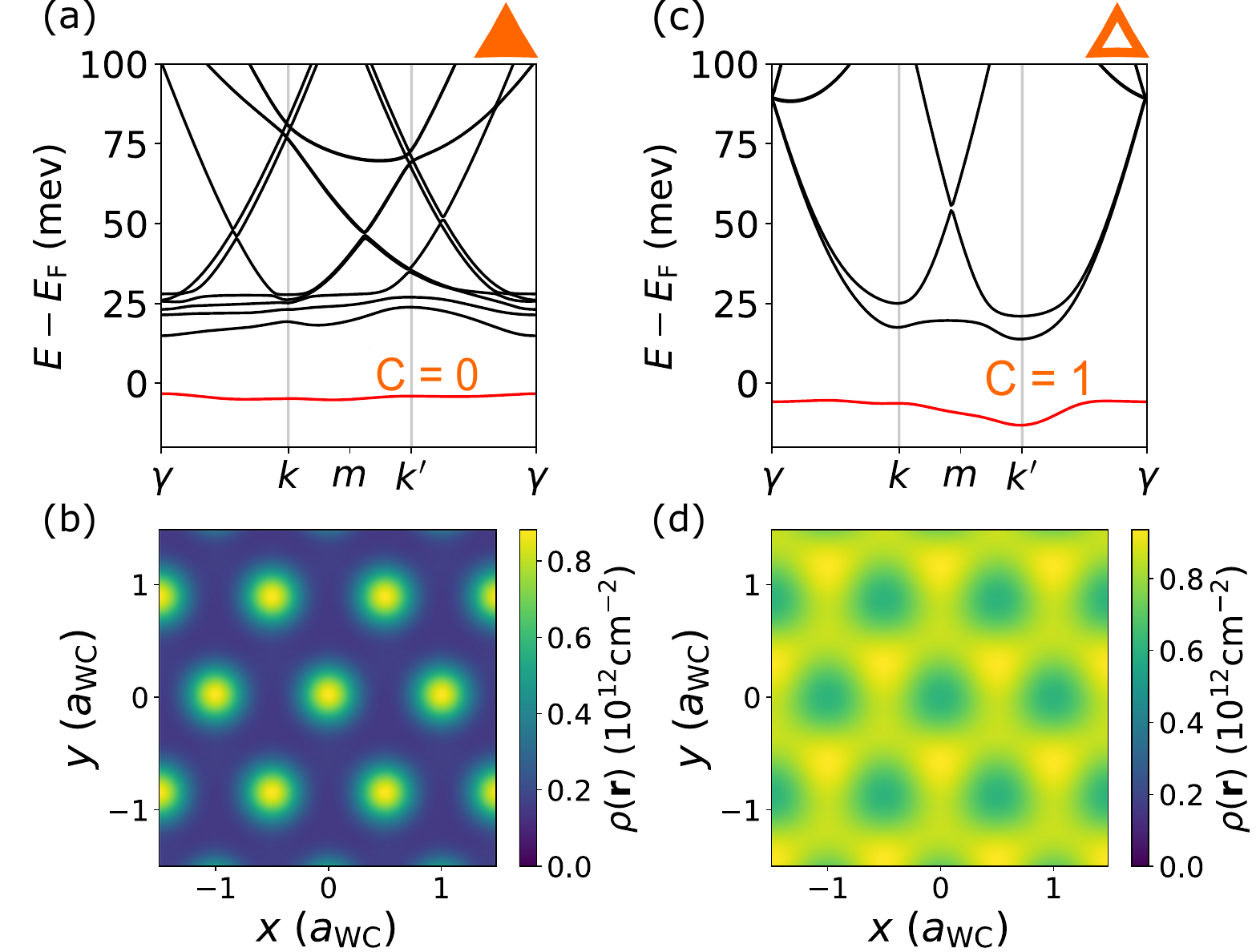} 
    \caption{Insulating WC states. 
    (a-b) The band structure and real space density distribution of the HF solution at $\delta_{\mathrm{D}} = -25\ \text{meV}$ and $n = 0.3\times 10^{12}\ \text{cm}^{-2}$. 
    (c-d) 
    Same plots but at $\delta_{\mathrm{D}} = -30\ \text{meV}$ and $n = 0.8\times 10^{12}\ \text{cm}^{-2}$.
    Red curves indicate the top valence band with Chern number $C$ indicated.
    }
    \label{fig:WC}
\end{figure}

\textit{\color{blue} Insulating Wigner crystal}---
At low electron densities, we find WC states.
Figs.~\ref{fig:WC}(a-b) respectively show the corresponding HF band structure and real space charge density distribution at $\delta_{\mathrm{D}} = -25\ \text{meV}$ and $n = 0.3\times 10^{12}\ \text{cm}^{-2}$. 
The HF band structure has a large gap between the lowest conduction and highest valence bands, while the density distribution in real space forms a triangle lattice.
The Chern number $C$ of the highest valence band [red curve in Fig.~\ref{fig:WC}(a)] vanishes.
A discussion of the Chern number evaluation and topological phase diagram can be found in Sec.~\ref{appsec:hf} in the SM \cite{supplemental_material}.

At higher electron density, the order parameter $|\rho({\mathbf{G}})|$ reduces without closing the direct gap in the HF band structure, signaling a 
 transition into a different insulating WC phase.
This behavior is in contrast to the 2D electron gas with a quadratic dispersion, where the WC states melt directly into the normal Fermi liquid as the density increases. 
Figs.~\ref{fig:WC}(c-d) respectively show the HF band structure and real space distribution at $\delta_{\mathrm{D}} = -30\ \text{meV}$ and a higher electron density $n = 0.8\times 10^{12}\ \text{cm}^{-2}$. 
The highest occupied HF band carries a non-zero Chern number $|C| = 1$.
By comparing Figs.~\ref{fig:WC}(b) and \ref{fig:WC}(d), the charge distribution of the crystalline lattice in the $|C| = 1$ case is more extended in real space than the $C = 0$ case 
and the peaks in charge density have shifted to a different $C_{3z}$-rotation center.

\begin{figure}[t]
    \centering                    
    \includegraphics[width=\linewidth]{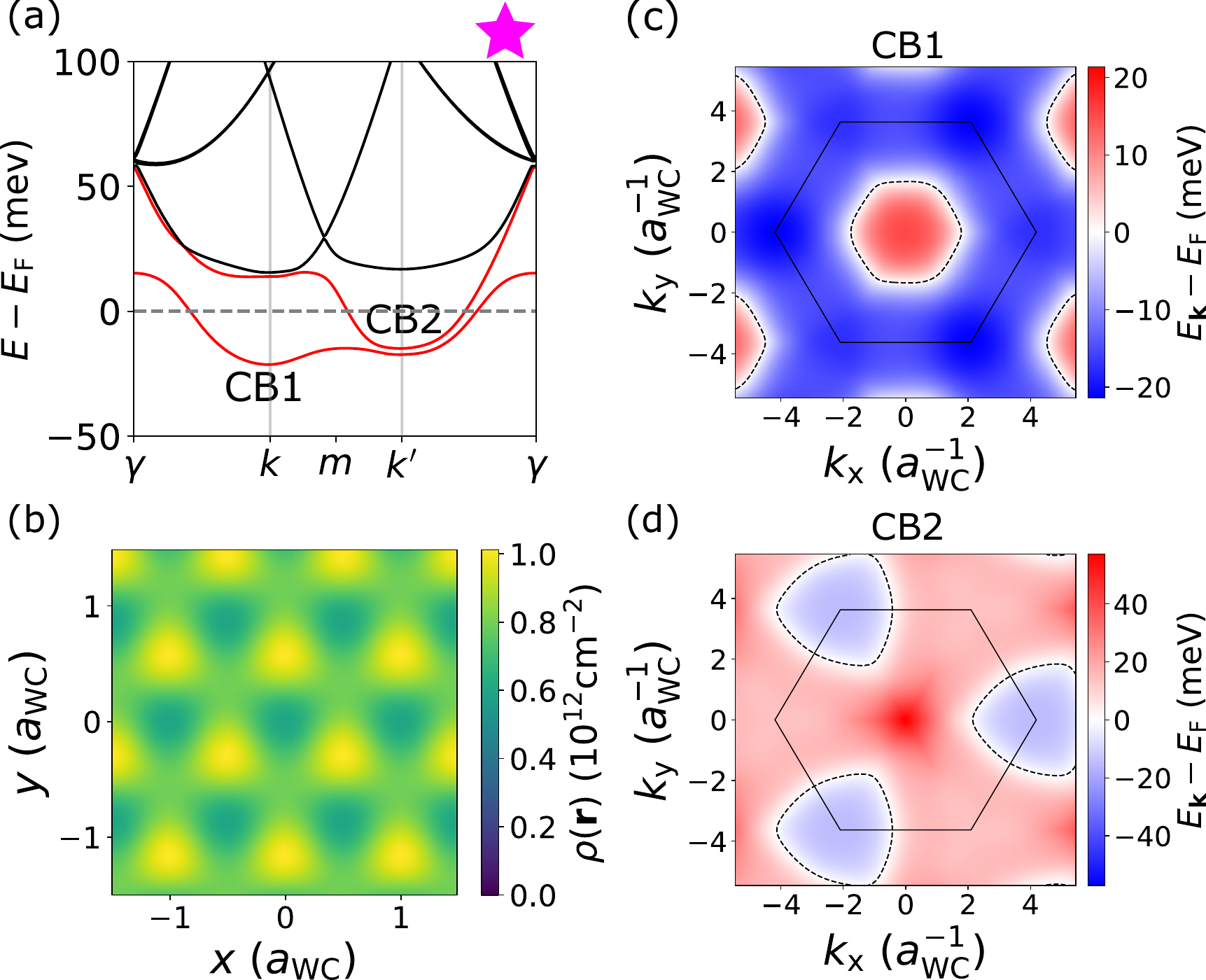} 
    \caption{
    Metallic WC state. 
    (a) HF band structure at $\delta_{\mathrm{D}} = -40\ \text{meV}$ and $n = 0.8\times 10^{12}\ \text{cm}^{-2}$.
    The two conduction bands near the Fermi level, denoted CB1 and CB2, are highlighted by red curves.
    (b) Real space density distribution. 
    (c-d) Dispersion of the first and second conduction bands (CB1 and CB2).
    The black solid line shows the Brillouin zone of this crystalline state, and the dashed curves show the Fermi surfaces, consisting of a hole 
    and 
    an electron pocket around the $\gamma$ and 
    $k'$ points, respectively.
    }           
    \label{fig:MWC}                              
\end{figure}

\textit{\color{blue} Metallic Wigner crystal}---
A larger interlayer potential elevates the degree of ``W''-shape in the bottom flat band, which
increases the effect of the coexisting electron and hole Fermi sheets.
Indeed, for interlayer potential in the range $-60\,{\rm meV} \lesssim \delta_{\rm D} \lesssim -40 \,\rm meV$, increasing the electron density to $n \sim 0.6 \times 10^{12}\,\rm cm^{-2}$
induces a transition from an insulating WC to a metallic one.
As shown in Fig.~\ref{fig:rho_G}(e), these metallic states retain a finite order parameter $|\rho({\mathbf{G}})|$.
To illustrate this, Fig.~\ref{fig:MWC} presents the HF band structure, Fermi surface and real space charge density distribution of a metallic WC state at $\delta_{\mathrm{D}} = -40\ \text{meV}$ and $n = 0.8\times 10^{12}\ \text{cm}^{-2}$. 
The real-space density distribution shown in Fig.~\ref{fig:MWC}(b) maintains a triangular crystalline structure, albeit with a more delocalized charge profile.
The dispersion of the two lowest conduction bands are shown in Figs.~\ref{fig:MWC}(c-d), 
with a hole Fermi pocket near the $\gamma$ point and an electron Fermi pocket near 
the $k'$ point. 
Since our HF is performed at commensurate filling, the two pockets necessarily enclose equal areas. 

\begin{figure}[t]
    \centering                    
    \includegraphics[width=\linewidth]{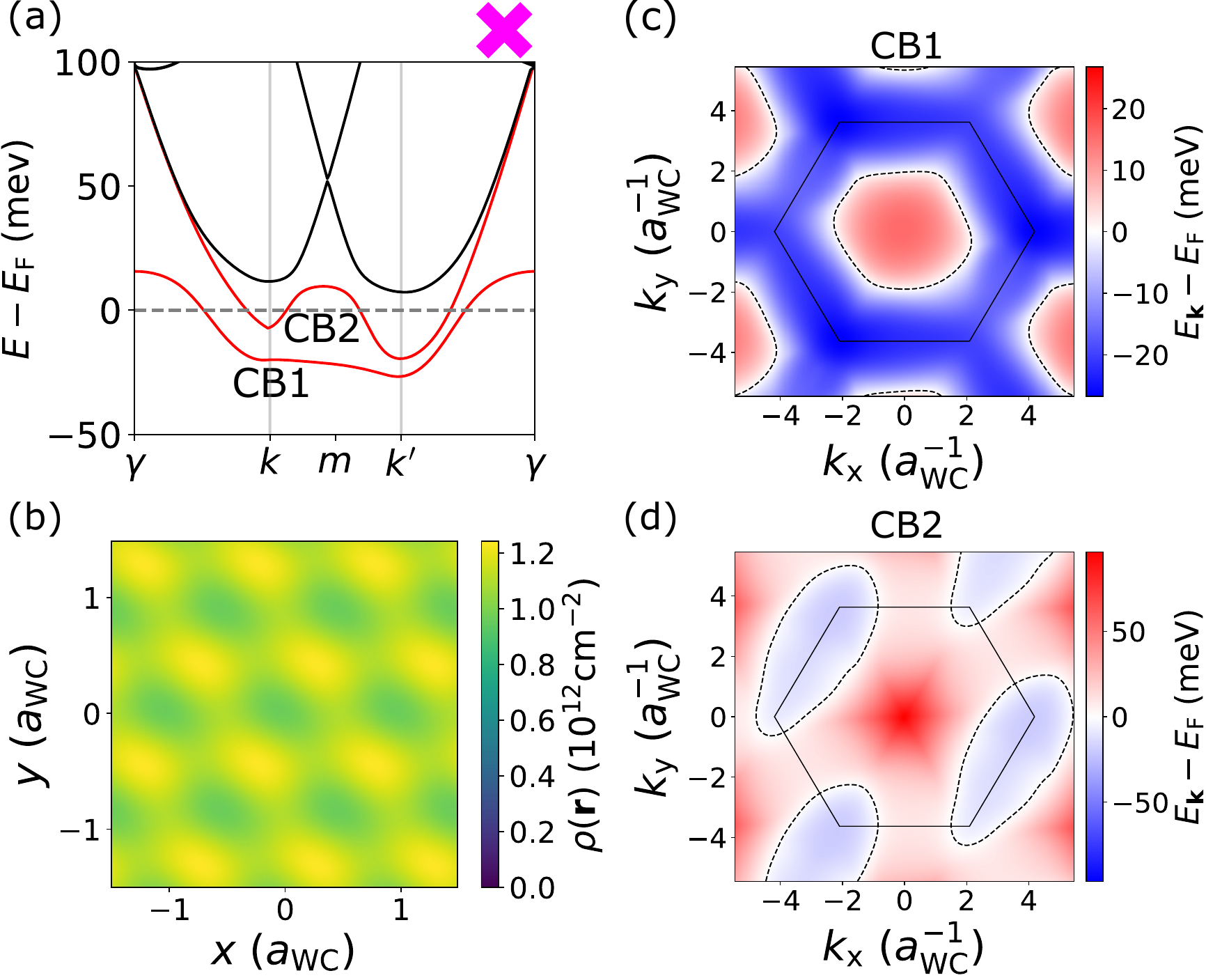} 
    \caption{Metallic WC state with $C_{3z}$-symmetry breaking.
    The model parameters are $\delta_{\rm D} = -45\,\text{meV}$ and $n = 1.1\times 10^{12}\,\text{cm}^{-2}$. 
    (a) HF band structure.
    (b) Real space density distribution.
    (c-d) Dispersion of the first and second conduction bands (CB1 and CB2). 
 The dashed curves show the Fermi surfaces, consisting of a hole pocket around $\gamma$ and an electron pocket that encloses a $k$-$m$-$k'$ segment.
    }
    \label{fig:MWC_C3z_breaking}                              
\end{figure}

\begin{figure}[t]
    \centering                    
    \includegraphics[width=0.6\linewidth]{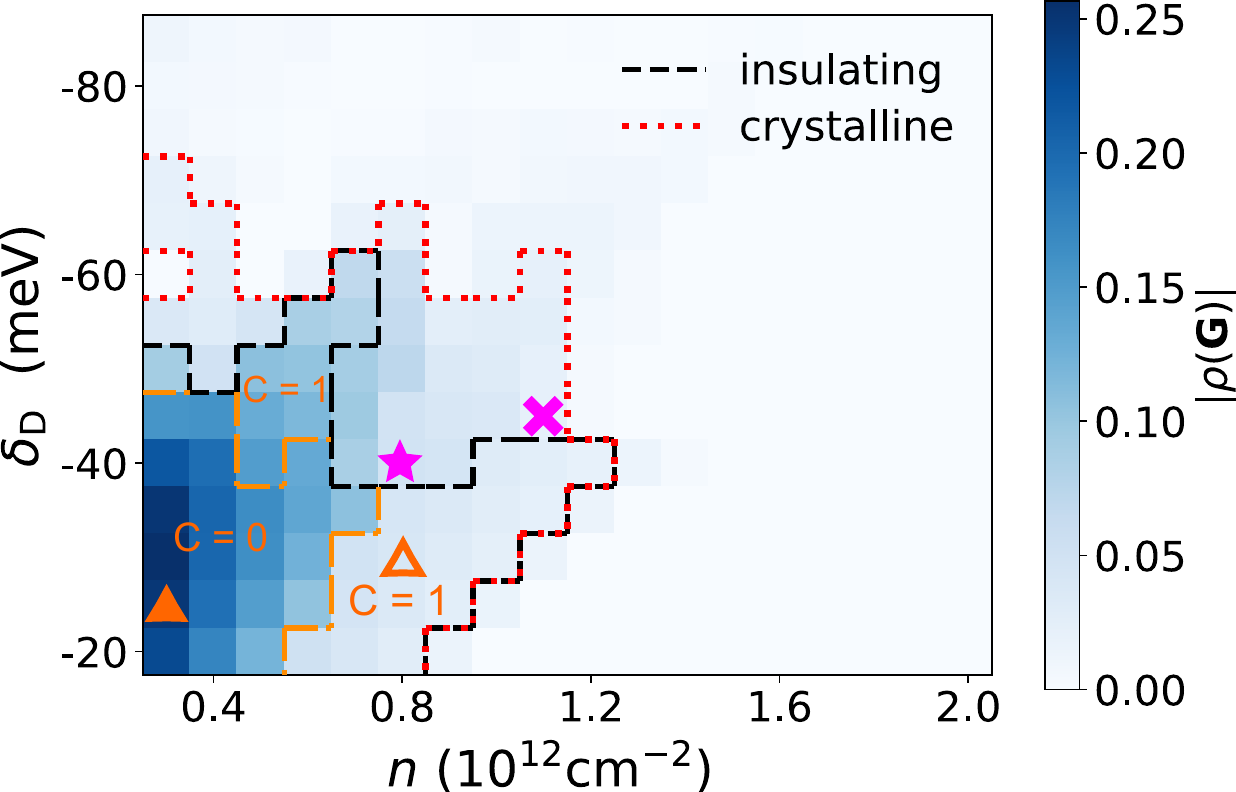} 
    \caption{
        Representative points shown in the main text Figs.~\ref{fig:WC},\ref{fig:MWC},\ref{fig:MWC_C3z_breaking} and in Fig~\ref{fig:RSMG continuum band} indicated. The orange dash-dotted line marks the boundary between $C = 0$ and $C = 1$ states in the insulating regime.
        Fig.~\ref{fig:M5N12_chern_number} presents the complete Chern number results.
    }
    \label{fig:M5N12_parameter_points}
\end{figure}

As the density of electrons increases further, the metallic WC states begin to melt. 
At the boundary between the crystalline and non-crystalline regions, 
a nematic state appears.
Fig.~\ref{fig:MWC_C3z_breaking} shows the HF band structure, Fermi surface and real space charge density distribution of a representative nematic state near the boundary at $\delta_{\mathrm{D}} = -45\ \text{meV}$ and $n = 1.1\times 10^{12}\ \text{cm}^{-2}$.
Its real space density distribution, shown in Fig.~\ref{fig:MWC_C3z_breaking}(b), breaks the
$C_{3z}$ rotational symmetry.
This feature is also reflected in the dispersion of the two conduction bands in Fig.~\ref{fig:MWC_C3z_breaking}(c-d).

\textit{\color{blue} Effective interacting lattice model} ---
The $C_{3z}$-broken phase revealed by our HF phase diagram motivates the construction of an effective real-space lattice model to elucidate how density-density interactions can generate nematic order with rotational symmetry breaking. 
Because this model is defined on the WC length scale, it is the natural analogue of the moir\'e-scale lattice models of tWSe$_2$, and thus serves to connect the two systems. It also provides a starting point for calculations beyond the HF approximation, including investigations of superconducting pairing.

\begin{figure}[t]
    \centering
    \includegraphics[width=\linewidth]{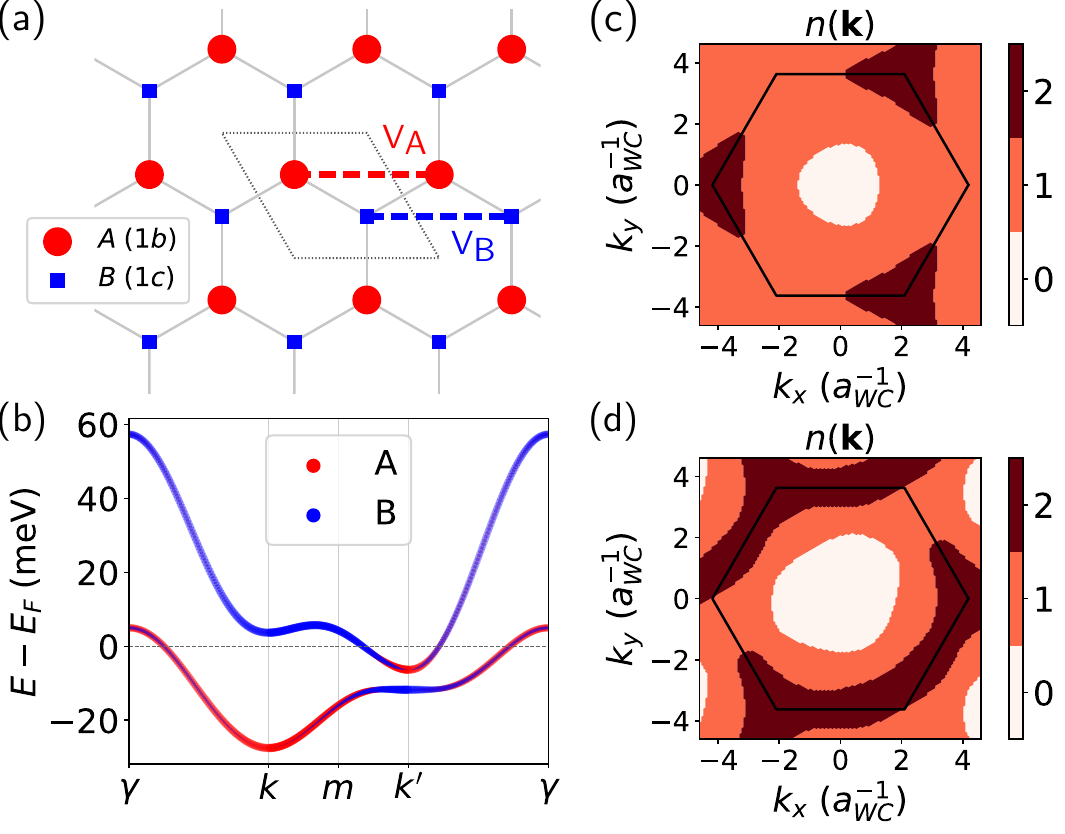}
    \caption{Effective $t$-$V$ model. (a) Schematic of the honeycomb lattice with  $A$ orbitals at the $1b$ Wyckoff position (red dots) and $B$ orbitals at the $1c$ Wyckoff position (blue squares).
    Marker sizes reflects the local real space density for each orbital. 
    Dotted lines mark the unit cell of the underlying triangular Bravais lattice, while dashed lines  denote density-density interaction bonds.  
    (b) Non-interacting band structure of the effective model, parametrized to reproduce the principal features of the two relevant HF bands at $\delta_\mathrm{D}=-40~\mathrm{meV}$ and $n=0.8\times10^{12}\mathrm{cm}^{-2}$. Red and blue indicate the $A$- and $B$-orbital weights. 
    (c, d) Momentum distribution  $n(\mathbf{k)}$ for $V_{A}=V_{B}=3~\mathrm{meV}$ and  $V_{A}=V_{B}=30~\mathrm{meV}$, respectively. 
    Black hexagons outline the first Brillouin zone.
    } 
    \label{fig:t-V}
\end{figure}

Guided by the lowest two conduction bands of the Hartree-Fock spectrum in the metallic and symmetry-preserving phase,
we construct a minimal spinless two-band lattice model, with orbitals $A$ and  $B$ residing at the $1b$ [$\boldsymbol{\tau}_A=(\mathbf{a}_1+2\mathbf{a}_2)/3$] and $1c$ [$\boldsymbol{\tau}_B=(2\mathbf{a}_1+\mathbf{a}_2)/3$] Wyckoff positions, as illustrated in Fig.~\ref{fig:t-V}(a).
The kinetic Hamiltonian $H_0$ is chosen to capture the principal dispersions and relative energy levels of CB1 and CB2 in Fig.~\ref{fig:MWC}(a),  with the resulting spectrum presented in Fig.~\ref{fig:t-V}(b).
The lower band carries a nonzero Chern number $|C|=1$, consistent with that of the corresponding band CB1. 
Both bands contribute near the Fermi level and their orbital character varies throughout the Brillouin zone, together motivating a two-orbital model. 
The explicit form of $H_0$ and its parametrization are described in the ``{\it End Matter}''.

Keeping the kinetic parameters fixed, we consider the full Hamiltonian $H=H_0+H_I$, with
\begin{equation}
    H_I=
    V_A \sum_{\langle\langle i,j\rangle\rangle}n_{iA}n_{jA}
    +V_B\sum_{\langle\langle i,j\rangle\rangle}n_{iB}n_{jB}.
    \label{eqn:interaction}
\end{equation}
Here $n_{i\alpha}=c_{i\alpha}^{\dagger}c_{i\alpha}$ denotes the occupation of orbital $\alpha=A,B$ in unit cell $i$.
The couplings $V_A$ and $V_B$
describe repulsive density-density interactions on next-nearest-neighbor bonds within sublattice $A$ and $B$,  respectively.
We perform self-consistent calculations at one electron 
per unit cell,
corresponding to half filling of the two-orbital Hilbert space while preserving primitive lattice translations.
For  small  interactions, $V_A=V_B=3~\mathrm{meV}$, 
the momentum occupation retains the threefold rotational symmetry, exhibiting a depleted region around $\gamma$ and double occupation near one set of symmetry related Brillouin zone corners [Fig.~\ref{fig:t-V}(c)]. 
Increasing the interactions to $V_A=V_B=30~\mathrm{meV}$ produces a pronounced 
anisotropy in the momentum occupation, as illustrated in Fig.~\ref{fig:t-V}(d).
The Fermi pockets shift and deform, and the resulting distribution no longer exhibits threefold rotation symmetry.
This spontaneous breaking of $C_{3z}$ symmetry, together with the preservation of primitive translations, identifies an electronic nematic state. 

As such, the proposed $t$-$V$ model sets the stage for studying superconductivity, which represents an alternative to nematicity. We expect the superconducting pairing 
to develop as the electrons reorganize to avoid the short-range repulsive interactions,
in parallel to what happens to the electrons of the spinful Hubbard model as they avoid the onsite repulsion.

\textit{\color{blue} Discussion}---
Several remarks are 
in order.
First,
we have emphasized 
how 
the ``W''-shaped dispersion in the flat bottom band
interplays with the long-range Coulomb interactions.
The crystalline order parameter reconstructs but preserves the coexisting
electron and hole Fermi sheets, leading to a metallic state.
Recent angle-resolved photoemission spectroscopy measurements in RMG provide direct evidence \cite{ming2026unpublished} for the ``W''-shaped dispersion.

Second, our metallic Wigner crystal states have important implications for the puzzling 
experimental results of the RMG. Because the Wigner crystallization reconstructs the 
{\it coexisting} electron and hole Fermi sheets
from the expectation of the unfolded continuum model 
(see Sec.~\ref{appsec:method} in the SM \cite{supplemental_material}), 
the Hall number can differ -- in magnitude and even in sign -- from the electron number.
In particular,
when the hole mobility is
considerably larger than the electron mobility as extracted for an R6G system \cite{Han2026Evidence}, our metallic Wigner crystal will 
have an unusual positive 
Hall number for an underlying electron filling, consistent 
with a recent experiment
\cite{Han2026Evidence}.
Furthermore, the Fermi surface reconstruction at the onset of the metallic Wigner crystal will lead to abrupt changes 
in the quantum oscillation frequencies, as also observed in recent experiments
\cite{Kalantre2026Fermiology, Zheng2026Chiral}.
These qualitative agreements set the stage
for more detailed comparisons in the future,
which will require the incorporation of the more realistic electronic structure.

Third,
the metallic WC serves as a candidate correlated normal state to nucleate 
superconductivity.
Our two-orbital $t$-$V$ model provides a basis to realize unconventional superconductivity.
In this regard, the effective $t$-$V$ model enables a parallel between RMG and tWSe$_2$ via their 
similar length scales and hybridized orbital structures \cite{xie2025kondo,li2025topological, li2026topology}. 

\textit{\color{blue} Conclusion}---
In summary, we have studied the correlated electronic structure of rhombohedral 
multilayer graphene 
in the dilute-carrier regime.
The periodicity at the atomic scale, captured by
the ``W''-shaped dispersion in the flat band,
interplays with the long-range Coulomb interaction and
enables the metallic WC state.
Near the phase boundary between the WC and disordered metallic regimes, we find such WC states to break the $C_{3z}$ symmetry, which
guides the construction of an effective two-orbital $t$-$V$ model. 
By identifying a
spontaneously generated long length scale in the strongly correlated metallic regime of RMG,
our work uncovers a hidden path to connect the mechanism for
superconductivity in 
RMG to its counterpart for the moir{\'e} systems such as tWSe$_2$.
As such, 
our findings suggest a new
route
towards the unification of superconductivity in a variety of dilute carrier systems.
Beyond the context of superconductivity, our work also sheds light on the key features that may give rise to new forms of
electronic crystals from an interplay between the long-range Coulomb repulsion
and an underlying atomic-scale periodicity,
such as in bulk materials settings \cite{Kim2022Purely}. 

\begin{acknowledgments}
\textit{Acknowledgments.}
We thank David Goldhaber-Gordon, Long Ju, Jia Li, Tomohiro Soejima, Yonglong Xie, and Ming Yi 
for useful discussions. 
This work has been supported in part by 
the U.S. Department of Energy, Office of Science, Basic Energy Sciences, under Award No. DE-SC0018197. J.C. and L.C. acknowledge support from the Air Force Office of Scientific Research under Grant No. FA9550-24-1-0222. J.C. also acknowledges support from the Flatiron Institute, a division of the Simons Foundation. Q.S. acknowledges the support and hospitality from Harvard University's Radcliffe Institute for Advanced Study (through a Radcliffe Fellowship) and Physics Department during the 2026–2027 academic year.
The majority of the computational calculations have been performed on the Shared University Grid at Rice funded by NSF under Grant No.~EIA-0216467, a partnership between Rice University, Sun Microsystems, and Sigma Solutions, Inc., the Big-Data Private-Cloud Research Cyberinfrastructure MRI-award funded by NSF under Grant No. CNS-1338099, and the Advanced Cyberinfrastructure Coordination Ecosystem: Services \& Support (ACCESS) by NSF under Grant No. DMR170109.  F.X., L.C., Q.S. and J.C. acknowledge the hospitality of the Aspen Center for Physics, which is supported by NSF grant No. PHY-2210452 and by a grant from the Simons Foundation (1161654, Troyer). N.N. acknowledges support from the JSPS Overseas Research Fellowship.
\end{acknowledgments}

\bibliography{ref}
\bibliographystyle{apsrev4-2}

\onecolumngrid
\begin{center}
\textbf{\large End Matter}
\end{center}
\twocolumngrid

{\color{blue} \it{Hartree-Fock method}}---
The Hamiltonian in Eq.~(\ref{eqn:continuum-interacting-hamiltonian}) contains a non-interacting and interacting part.
The non-interacting Hamiltonian $H_{\rm R5G}$ describes non-interacting long-wavelength band structure of a quarter metal with fully polarized spin and valley, within a ten-band continuum model.
It contains five copies of Dirac fermions with Fermi velocity $v_0 = 5904 \, \rm meV \cdot \text{\AA}$, representing the $K$-valley band structure of each graphene layer, as well as nearest-neighbor interlayer hopping $t_1 = 340{\rm \,meV}$ and next-nearest-neighbor interlayer hopping $t_2 = -10\,\rm meV$ \cite{Zhang2010Band, Dresselhaus1981Intercalation}.
Applying a vertical displacement field shifts the on-site energy of each layer by $\delta_{\rm D}$, generating the ``W''-shaped band bottom discussed in the main text.
The non-interacting Hamiltonian also inherits the $C_{3z}$ rotation and the $TM_x$ symmetry from the underlying rhombohedral stacking, where $T$ and $M_x$ denote the time-reversal and mirror operations with respect to the $yz$ plane, respectively.
The non-interacting Hamiltonian is written explicitly in  Sec.~\ref{appsec:r5g} of the SM \cite{supplemental_material}.

The electron-electron interaction we consider is a double-gate screened Coulomb potential $V(\mathbf{q}) = {e^2\tanh{(qd)}}\big{/}({2\epsilon_0\epsilon_r q})$. We choose the gate distance $d = 20\, \text{nm}$ and the relative permittivity $\epsilon_r = 7$.

To study the possible crystalline states, we perform self-consistent Hartree-Fock (HF) calculations under the assumption of full spin and valley polarization.
Since the Hamiltonian Eq.~(\ref{eqn:continuum-interacting-hamiltonian}) is defined in continuum space without a lattice structure, we employ a triangular lattice {\it ansatz} at commensurate filling with exactly one electron per unit cell.
For a given density of electrons $n$, the lattice constant is given by $a_{\rm WC} = \sqrt{{2}/({\sqrt{3}n})}$.
Throughout this work, keep the lattice vectors $\mathbf{a}_1 = (1,0)a_{\rm WC}$ and $\mathbf{a}_2 = \left(-{1}/{2},{\sqrt{3}}/{2}\right)a_{\rm WC}$ fixed.
Therefore, any wave vector $\vp$ can always be represented as $\vp = \vk + \mathbf{G}$, where $\vk$ is a Bloch wave vector within the first Brillouin zone and $\mathbf{G}$ is a reciprocal vector.
The Hartree-Fock Hamiltonian is then derived separately within each sector labeled by the quantum number $\vk$. 
Further details about the HF calculation can be found in Sec.~\ref{appsec:hf-topology} in the SM \cite{supplemental_material}.

In Sec.~\ref{appsec:hf-stability} of the SM \cite{supplemental_material}, we describe the free energy and demonstrates the stability of our HF solution. Beyond our specific calculations, we expect the general theme, viz. an emergent length scale spontaneously generated by the long-range Coulomb interactions bridges the low-energy physics of the metallic dilute-electron RMGs to its counterpart in tWSe$_2$,
to be robust.

{\color{blue} \it{Phase diagram}}---
The phase diagram is characterized by the crystalline order parameter
$|\rho(\mathbf{G})|$, defined as:
\begin{equation}
    \rho(\mathbf{G}) = \frac{1}{N_{\mathbf{k}}}\sum_{\mathbf{k}\in\text{BZ}}\sum_{\mathbf{Q},\alpha}\braket{a^{\dagger}_{\mathbf{k}+\mathbf{Q},\alpha}a_{\mathbf{k}+\mathbf{Q+G},\alpha}}\,,
\end{equation}
where $\mathbf{Q}, \mathbf{G}$ stand for reciprocal vectors according to the lattice basis vectors $\mathbf{a}_1$, $\mathbf{a}_2$, and $\alpha$ is a composite index combining the layer and sublattice degrees of freedom.
$|\rho(\mathbf{G})|$ determines the Bragg term of the structure factor in our HF solution.
A detailed discussion about $|\rho(\mathbf{G})|$ can be found in Sec.~\ref{appsec:hf} of the SM~\cite{supplemental_material}. 
The line cuts, shown for two representative values of $\delta_{\rm D}$ in Figs.~\ref{fig:rho_G}(d-e),
exhibit a step-like suppression
with increasing electron density, before it decreases further towards the vanishing limit.
eventually vanishing.
In addition to the crystalline boundary (red dotted line) and the metal-insulator boundary (black dashed line), there is also a boundary marking a topological transition. 
The latter is
shown in Fig.~\ref{fig:M5N12_parameter_points}. 
As addressed in Sec.~\ref{appsec:hf-topology} of the SM \cite{supplemental_material}, this transition is first order.

{\color{blue} \it{$t$-$V$ model}}---
We consider a spinless two-band Hamiltonian defined on a honeycomb lattice, 
with two local orbitals $A$ and $B$ centered at the $1b$ [$\boldsymbol{\tau}_A=(\mathbf{a}_1+2\mathbf{a}_2)/3$] and $1c$ [$\boldsymbol{\tau}_B=(2\mathbf{a}_1+\mathbf{a}_2)/3$] Wyckoff positions, respectively.
The kinetic part takes the form:
\begin{align}
    H_0=&
    \sum_{\mathbf R,\alpha}
    \varepsilon_\alpha n_{\mathbf R\alpha}
    +\sum_{\mathbf R,\ell}
    \left(t_{AB}c_{\mathbf RA}^{\dagger}
    c_{\mathbf R+\boldsymbol{\eta}_\ell,B}
    +\mathrm{h.c.}
    \right)\nonumber\\
    &+\sum_{\mathbf R,\ell,\alpha}
    \left(t_\alpha c_{\mathbf R\alpha}^{\dagger}
    c_{\mathbf R+\boldsymbol{\delta}_\ell,\alpha}
    +\mathrm{h.c.}
    \right) \, ,
\end{align}
where $c_{\mathbf{R}\alpha}^{\dagger}$ creates a fermion at site $\mathbf{R}+\boldsymbol{\tau}_\alpha $,  $n_{\mathbf{R}\alpha}=c_{\mathbf{R}\alpha}^{\dagger}c_{\mathbf{R}\alpha}$, and the bond vectors are defined by   
\begin{align}
    (\boldsymbol{\eta}_1,\boldsymbol{\eta}_2,\boldsymbol{\eta}_3)
    &=(\mathbf 0,-\mathbf a_1,\mathbf a_2),
    \nonumber\\
    (\boldsymbol{\delta}_1,\boldsymbol{\delta}_2,\boldsymbol{\delta}_3)
    &=(\mathbf a_1,\mathbf a_2,-\mathbf a_1-\mathbf a_2)\, .
\end{align}
The amplitudes $t_{AB}$ describes the nearest-neighbor hopping between opposite sublattices, while $t_\alpha$ describes next-nearest neighbor hopping within sublattice $\alpha\in(A, B)$. 
The orbital energies and hoppings are chosen  to 
qualitatively match the lowest two conduction bands of the $C_{3z}$ symetric metallic HF solution shown in Fig.~\ref{fig:MWC}(a):
\begin{align}
&\varepsilon_A=0~\mathrm{meV}, \quad t_A = 4e^{-i0.53\pi}~\mathrm{meV}, \nonumber\\
&\varepsilon_B=20~\mathrm{meV}, \quad t_B = 6e^{i0.25\pi}~\mathrm{meV}, \nonumber\\
&t_{AB} = 6.6~\mathrm{meV}. 
    \label{eqn:parameters}
\end{align}
For this set of parameters, the real space charge densities are $n_{iA}=0.78$ and $n_{iB}=0.22$, forming an effective triangular lattice.
In addition to capturing the principal band dispersions,
the lower band reproduces the rotation eigenvalues $(\omega,\omega^*,\omega^*)$
of the lowest reference HF conduction band at $(\gamma,k,k')$,
respectively. 
These symmetry eigenvalues are consistent with the band's non-zero Chern number $|C|=1$.

With the interaction terms defined in Eq.~(\ref{eqn:interaction}), 
self-consistent calculations are performed in momentum space on a $96\times96$ grid at one fermion per site. 
To allow rotational symmetry breaking while preserving primitive translational invariance, we restrict the one-body density matrix to $\langle c_{\mathbf{k}\alpha}^\dagger c_{\mathbf{k}'\beta}\rangle=\delta_{\mathbf{k},\mathbf{k}'}\rho_{\alpha\beta}(\mathbf{\mathbf{k}})$, without imposing $C_{3z}$ symmetry.

{\color{blue} \it{Scope of this work}}---
We close 
this 
{\it End Matter}
with two further points concerning the theoretical scope of our 
study.
To begin with,
rotational symmetry breaking also appears in
the nematic insulating WC 
of R4G studied in Ref.~\cite{Ge2026Nematic}.
However, in addition to the metallic nature of our solution,
the $C_{3z}$ breaking in our calculation 
retains the translational symmetry of the triangular lattice,
instead of 
deforming the lattice vectors;
the rotational symmetry breaking is realized in the electronic density distribution and reconstructed bands. 
This procedure is important in the construction of the effective $t$-$V$
description.

Furthermore, 
our $t$-$V$ model examines electronic ordering on a fixed WC background. 
It does not explicitly retain the collective displacements of the electronic crystal, such as the WC phonons \cite{Crandall1973Collective,Platzman1974Phase,Bonsall1977Some}.
Although the model allows empty and doubly occupied sites, identifying these configurations with physical vacancies and interstitials requires accounting for the accompanying lattice relaxation and defect energetics \cite{Kim2024Dynamical, Matus2026theory}. 
The added effect of these collective excitations and defect relaxations 
represents a promising direction for future research.

\clearpage

\onecolumngrid
\beginsupplement
\section*{Supplemental Material}
\setcounter{secnumdepth}{3}

\tableofcontents

\section{The model and Hartree-Fock approach}\label{appsec:method}

In this section, we briefly review the low energy effective Hamiltonian for R5G systems, the form of the interacting Hamiltonian, and the Hartree-Fock mean field theory.

\subsection{Effective Hamiltonian of single-layer rhombohedral stacked multi-layer graphene}\label{appsec:r5g}

We first consider the band structure of R5G without electronic correlations.
The effective Hamiltonian of R5G system in a single valley has the following form:
\begin{equation}
    h_{\rm R5G}(\vk) = \left(
        \begin{matrix}
            h_D(\vk) + 2\delta_{\rm D} & V & W & \mathbb{0} & \mathbb{0}  \\
            V^\dagger & h_D(\vk) + \delta_{\rm D} & V & W & \mathbb{0} \\
            W^\dagger & V^\dagger & h_D(\vk) & V & W \\
            \mathbb{0} & W^\dagger & V^\dagger & h_D(\vk) - \delta_{\rm D}& V\\
            \mathbb{0} & \mathbb{0} & W^\dagger & V^\dagger & h_D(\vk) - 2\delta_{\rm D}
        \end{matrix}
    \right)\,,
\end{equation}
in which $h_D(\vk)$ is the Dirac Hamiltonian, which effectively describes the low energy dispersion of graphene near its $K$ point of the single layer Brillouin zone;
$V$ and $W$ are the nearest and next-nearest neighbor interlayer hopping matrices;
and $\delta_{\rm D}$ is the interlayer potential, induced by the displacement field perpendicular to the graphene sheets.
The concrete form of the three matrices $h_D(\vk), V$ and $W$ are given by:
\begin{equation}
    h_D(\vk) = \left(
        \begin{matrix}
            0 & v_0 (k_x - i k_y) \\
            v_0 (k_x + ik_y) & 0
        \end{matrix}
    \right)\,, ~~ V = \left(
        \begin{matrix}
            0 & 0\\
            t_1 & 0
        \end{matrix}
    \right)\,, ~~ W = \left(
        \begin{matrix}
            0 & t_2 \\
            0 & 0
        \end{matrix}
    \right)\,.
\end{equation}
where the Fermi velocity of single layer graphene $v_0 = 3t_0 a_G/2$ is determined by the nearest neighbor hopping $t_0$ and graphene carbon-carbon distance $a_G$.
In this work, we choose the Fermi velocity as $v_0 = 5904 \rm meV\cdot \text{\AA}$.
The nearest-neighbor and next-nearest-neighbor interlayer hoppings are $t_1 = 340.0\,\rm meV$ and $t_2 = -10.0 \,\rm meV$, respectively \cite{Zhang2010Band, Dresselhaus1981Intercalation}.
Note that this Hamiltonian describes the long wavelength behavior of the Bloch electrons, and is not valid for describing the short distance physics on the atomic scale.
Hence, in the following, we will refer to this model as the ``continuum model.''

\begin{figure}[htbp]
    \centering                
    \includegraphics[width=0.6\linewidth]{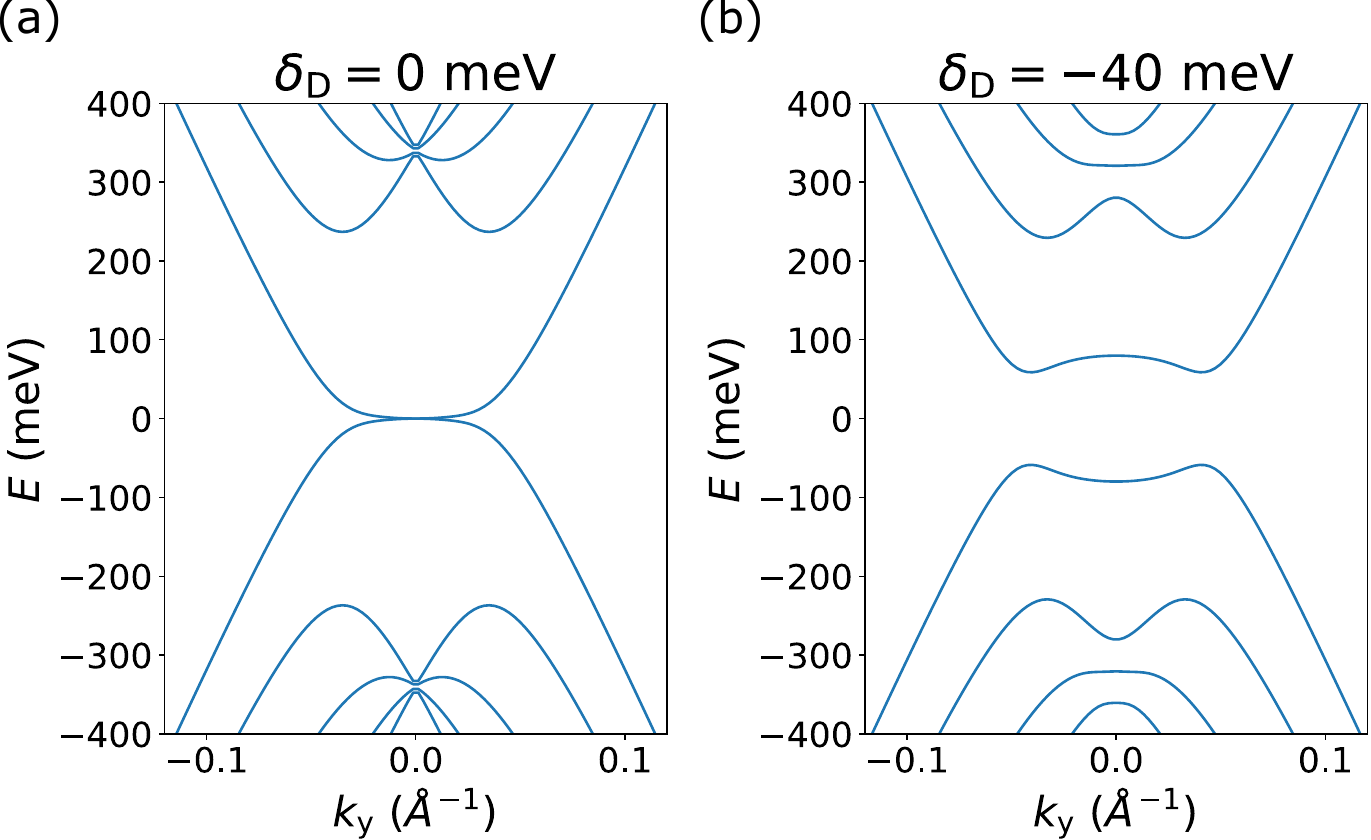} 
    \caption{
        Band structure of single valley Hamiltonian of R5G under interlayer potential (a)  $\delta_{\rm D} = 0 \,\rm meV$ and (b) $\delta_{\rm D} = -40 \,\rm meV$.
    }
    \label{fig:10_band}                         
\end{figure}

\begin{figure}[t]
    \centering                    
    \includegraphics[width=1\linewidth]{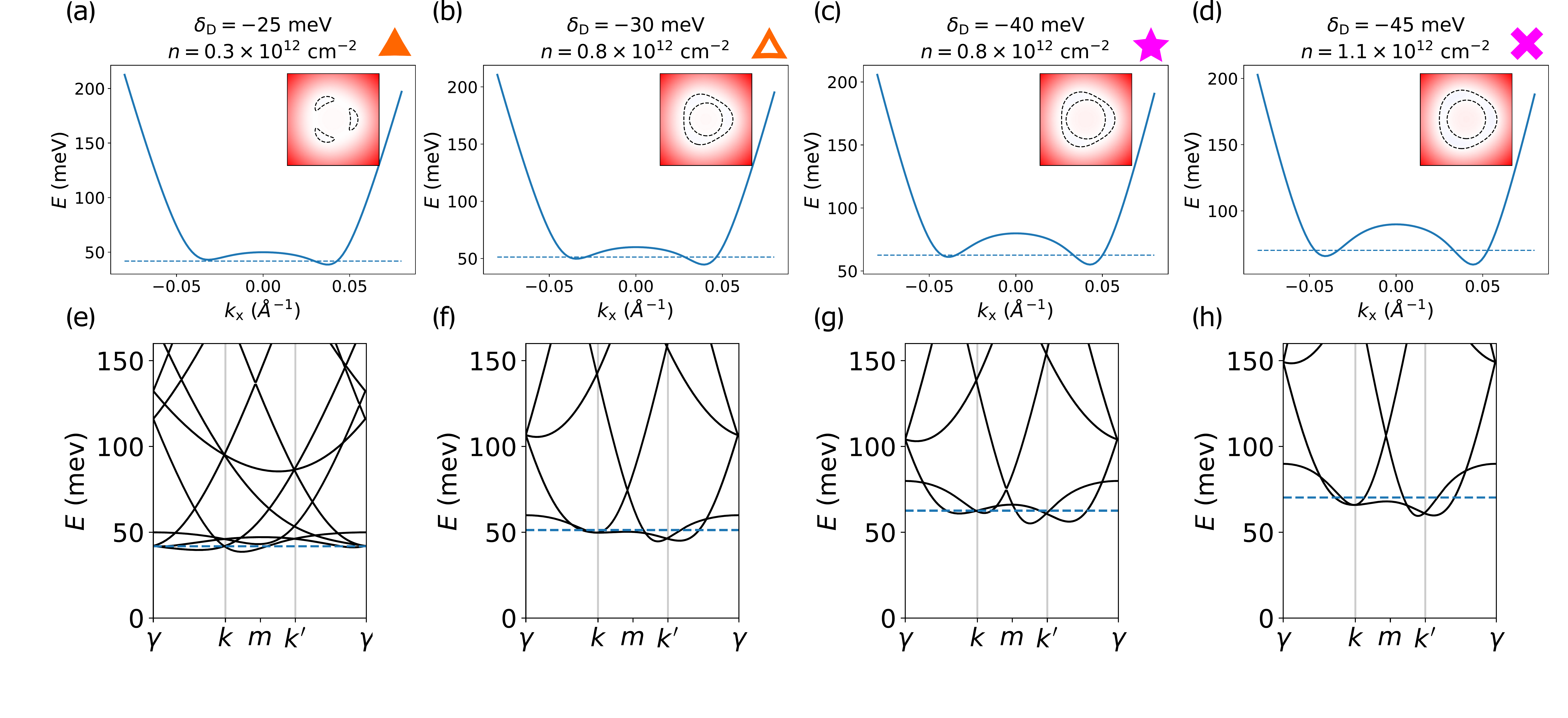} 
    \caption{
        Upper panels: the band dispersion of the first band above charge neutrality at different displacement fields $\delta_{\rm D}$ labeled at top of the panels;
        blue dashed lines represent the Fermi level at different electron densities $n$, also labeled at the top;
        the insets show the corresponding Fermi surface geometries.
        Lower panels: the band structure plots of the continuum model, folded into the first Brillouin zone.
        The model parameters correspond to those used in the main text Figs.~\ref{fig:WC}, \ref{fig:MWC} and \ref{fig:MWC_C3z_breaking} in the main text.
    }
    \label{fig:RSMG continuum band}
\end{figure}

By diagonalizing $h_{\rm R5G}(\vk)$ with different interlayer potential values, we obtain the non-interacting band structure of R5G around the $K$ point.
Fig.~\ref{fig:10_band} shows the dispersion of all ten bands with interlayer potential $\delta_{\rm D} = 0\, \rm meV$ and $\delta_{\rm D} = -40 \,\rm meV$, respectively. 
In the absence of the interlayer potential, there is a band touching point between the two bands near charge neutrality, accompanied by an extremely flat band dispersion.
When the interlayer potential is applied, a gap opens between these bands, and the lowest conduction band can develop a ``W''-shape dispersion near its bottom band edge.
The upper panels in Fig.~\ref{fig:RSMG continuum band} present the continuum model band structure for the parameter sets used in Figs.~\ref{fig:WC}, \ref{fig:MWC} and \ref{fig:MWC_C3z_breaking} in the main text.
Both the insulating and metallic WC states discussed in the main text exist when the Fermi level sits within the ``W''-shaped band bottom region.

\subsection{Interacting Hamiltonian and Hartree-Fock method}\label{eqn:int-hf}

In this work, we consider a screened Coulomb interaction.
The interacting Hamiltonian has the following form:
\begin{equation}
    H = \sum_{\mathbf{k}}\sum_{\alpha\beta}[h_{\text{R5G}}(\mathbf{k})]_{\alpha\beta} a^{\dagger}_{\mathbf{k},\alpha} a_{\mathbf{k},\beta} + \frac{1}{2A}\sum_{\mathbf{k,k',q}}\sum_{\alpha\beta}V(\mathbf{q}) a^{\dagger}_{\mathbf{k+q},\alpha} a^{\dagger}_{\mathbf{k'-q},\beta} a_{\mathbf{k'},\beta} a_{\mathbf{k},\alpha} \,,
    \label{eqn:ham}
\end{equation}
where $\alpha, \beta$ indicate the sublattice indices of single layer graphene;
$V(\mathbf{q}) = \frac{e^2\tanh{(q d})}{2\epsilon_0\epsilon_r q}$ is the Fourier transform of the double-gate screened Coulomb potential, $d$ is the gate distance and $\epsilon_r$ is the relative permittivity.
The summation over $\vk, \vk'$ and $\vq$ run over all wave vectors allowed by the boundary conditions.

To study possible crystallized phase in R5G, we implement a restricted Hartree-Fock mean field calculation.
We assume spontaneous breaking of continuous translation symmetry via the formation of a triangular crystal lattice with primitive vectors
$\mathbf{a}_1 = \left(1,0 \right)a_{\rm WC},\,\mathbf{a}_2 = (-1/2,\sqrt{3}/2)a_{\rm WC}$, where $a_{\rm WC}$ is the WC lattice constant. 
At commensurate filling (one electron per unit cell),  $a_{\rm WC}$  is fixed by the electron density $n$ via 
\begin{equation}
    a_{\rm WC} = \sqrt{{2}/({\sqrt{3}n})}\,.
\end{equation}
The corresponding reciprocal lattice vectors are
$\mathbf{b}_1 = (\sqrt{3}/2,1/2)b_{\mathrm{WC}},\,\mathbf{b}_2 = \left(0,1 \right) b_{\mathrm{WC}}$, with $b_{\mathrm{WC}} = 4\pi/(\sqrt{3}a_{\rm WC})$. 
We can fold any wave vector $\vp$ into the first Brillouin zone (BZ) as $\vp = \vk + \mathbf{Q}$, where $\mathbf{k}$ is in the first Brillouin zone, and $\mathbf{Q} = m_1\mathbf{b}_1 + m_2\mathbf{b}_2$, $m_1,m_2 \in \mathbb{Z}$, is a reciprocal vector.
Treating $\mathbf{Q}$ as an ``internal'' index,
we re-index the fermionic operator $a^{\dagger}_{\mathbf{k}+\mathbf{Q},\alpha}$ as $a^{\dagger}_{\mathbf{k}, \mathbf{Q},\alpha}$.
In this plane wave basis, the Hamiltonian in Eq.~(\ref{eqn:ham}) can be rewritten as:
\begin{align}
    H =& \sum_{\mathbf{k}\in\text{BZ}}\sum_{\mathbf{Q,Q'}}\sum_{\alpha\beta} [h_{\rm R5G}(\vk + \mathbf{Q})]_{\alpha\beta} \delta_{\mathbf{Q}, \mathbf{Q}'} a^{\dagger}_{\mathbf{k,Q,\alpha}}a_{\mathbf{k,Q',\beta}} \nonumber \\
    & +\frac{1}{2A}\sum_{\mathbf{k,k',q}\in\text{BZ}}\sum_{\mathbf{Q,Q',G}}\sum_{\alpha\beta}V(\mathbf{q+G})a^{\dagger}_{\mathbf{k+q,Q+G,\alpha}}a^{\dagger}_{\mathbf{k'-q,Q'-G},\beta}a_{\mathbf{k',Q',\beta}}a_{\mathbf{k,Q,\alpha}} \,.
\end{align}

Hartree-Fock is a variational method using Slater determinants  $\ket{\psi}$ as the ansatz.
Each state is characterized by the one-particle density matrix $\rho_{\mathbf{Q\alpha,Q'\beta}}(\mathbf{k}) = \bra{\psi}{a^{\dagger}_{\mathbf{k,Q,\alpha}}a_{\mathbf{k,Q',\beta}}}\ket{\psi}$.
To eliminate double counting the interaction, we perform the subtraction:
\begin{equation}
    \delta\rho_{\mathbf{Q\alpha,Q'\beta}}(\mathbf{k}) = \rho_{\mathbf{Q\alpha,Q'\beta}}(\mathbf{k})-[\rho_{\text{CN}}]_{\mathbf{Q\alpha,Q'\beta}}(\mathbf{k}) \,,
\end{equation}
where $\rho_{\text{CN}}$ is the density matrix of the non-interacting ground state at charge neutrality.
As such, it is ensured that the background interactions at charge neutrality are already incorporated into the non-interacting continuum model.
Using Wick's theorem, the mean-field Hamiltonian, which depends on the density matrix $\rho$, has the following form:
\begin{align}
    [h_{\text{MF}} (\vk)]_{\mathbf{Q}\alpha, \mathbf{Q}'\beta} &= [h_{\rm R5G}(\vk+\mathbf{Q})]_{\alpha\beta}\delta_{\mathbf{Q}\mathbf{Q}'} + [h^{(H)}(\mathbf{k})+ h^{(F)}(\mathbf{k})]_{\mathbf{Q\alpha,Q'\beta}} \,, \label{eqn:mf-ham} \\
    h^{(H)}_{\mathbf{Q\alpha,Q'\beta}}(\vk, \delta\rho) &= \frac{1}{A}\sum_{\mathbf{G}}\left(\sum_{\mathbf{k'}\in\text{BZ}}\sum_{\mathbf{Q''}}\sum_{\alpha'}V(\mathbf{-G})\delta\rho_{\mathbf{(Q''+G)\alpha',Q''\alpha'}}(\mathbf{k'})\right)\delta_{\alpha\beta}\delta_{\mathbf{Q',Q+G}} \,, \\ 
    h^{(F)}_{\mathbf{Q\alpha,Q'\beta}}(\vk, \delta\rho) &= -\frac{1}{A}\sum_{\mathbf{G}}\left(\sum_{\mathbf{k'}\in\text{BZ}}\sum_{\mathbf{Q''}}V(\mathbf{k'-k+Q''-Q})\delta\rho_{\mathbf{(Q''+G)\beta,Q''\alpha}}(\mathbf{k'})\right)\delta_{\mathbf{Q',Q+G}} \,.
\end{align}
Diagonalizing this mean-field Hamiltonian, we obtain the HF orbital energies $E_n$ and single-particle orbital wave functions $u$:
\begin{equation}\label{eqn:hf-eigen-eqn}
    \sum_{\mathbf{Q'\beta}}[h_{\rm HF}(\vk)]_{\mathbf{Q\alpha,Q'\beta}}u_{\mathbf{Q'\beta},n}(\vk) = E_n(\vk)u_{\mathbf{Q\alpha},n}(\vk)\,.
\end{equation}
The density matrix $\rho$ can therefore be determined by these single-particle wave functions as follow:
\begin{equation}\label{eqn:density-matrix}
    \rho_{\mathbf{Q}\alpha, \mathbf{Q}'\beta}(\vk) = \sum_n u^*_{\mathbf{Q}\alpha, n}(\vk) u_{\mathbf{Q}'\beta, n}(\vk)\,,
\end{equation}
where the band index $n$ runs over all occupied states.
Eqs.~(\ref{eqn:mf-ham}-\ref{eqn:density-matrix}) define the HF framework of the R5G system, allowing for continuous translation symmetry breaking that satisfies commensurate filling.

These equations can be solved in a self consistent manner with a given initial condition.
After a solution is obtained, the energy of the converged state is given by:
\begin{equation}
    E_{\text{HF}} = \bra{\psi}H\ket{\psi} = \sum_{\mathbf{k}\in\text{BZ}}\sum_{\mathbf{Q,Q'}}\sum_{\alpha\beta}\left[\tilde{h}_{\text{R5G}}(\mathbf{k}) + \frac{1}{2}h^{(H)}(\mathbf{k}, \delta\rho)+\frac{1}{2}h^{(F)}(\mathbf{k}, \delta\rho)\right]\delta\rho_{\mathbf{Q\alpha,Q'\beta}}(\mathbf{k})\,.
\end{equation}

\begin{figure}[htbp]
    \centering                    
    \includegraphics[width=0.6\linewidth]{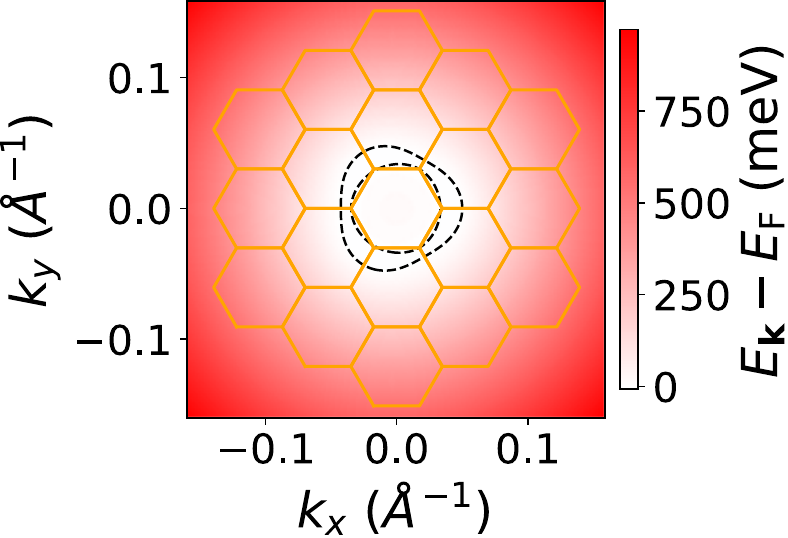} 
    \caption{Brillouin zone folding and Fermi surfaces with representative model parameters. 
    Color indicates the energy of the first conduction band under interlayer potential $\delta_{\mathrm{D}} = - 40 \ \text{meV}$ [see the inset of Fig.~\ref{fig:RSMG continuum band}(c)].
    The black dashed line labels the Fermi surface at electron density $n = 0.8 \times 10^{12} \ \text{cm}^{-2}$, and the orange hexagons represent the Brillouin zones we include in the HF calculation. 
    }
    \label{fig:BZ_folding}                              
\end{figure}

In this work, we consider $N_{\mathbf{Q}} = 19$ different Brillouin zone points, which constitutes two ``shells'', as shown in Fig.~\ref{fig:BZ_folding}.
Taking $N_{\mathbf{Q}} = 19$ is sufficient to capture the ``W''-shaped feature of the conduction band bottom, which is an essential feature for the interacting physics.
We consider an $N_{\mathbf{k}} = 12\times 12$ momentum mesh in each Brillouin zone.
The non-interacting continuum model band structure folded into the WC Brillouin zone is shown in the lower panels of Fig.~\ref{fig:RSMG continuum band} for the parameter values considered in the main text.

We choose seven randomly generated initial conditions and perform the self-consistent cycle for each point in the phase diagram, and the solution with the lowest converged energy are presented.

\subsection{Non-self-consistent calculation for the HF bands along high symmetry lines}
\label{appsec:nonselfconsistent}

Our $12\times12$ momentum grid in the first Brillouin zone is not enough to produce a smooth band along high symmetry lines. 
Nonetheless, by using the density matrix obtained from the $12 \times 12$ grid, we are able to construct $h_{\rm MF}(\vk)$ for other $\mathbf{k}$ points in the Brillouin zone: 
\begin{align}
    h_{\rm MF}(\mathbf{k}) &= \tilde{h}_{\text{R5G}}(\mathbf{k}) + \tilde{h}^{(H)}(\mathbf{k}) + \tilde{h}^{(F)}(\mathbf{k})
    \tilde{h}^{(H)}_{\mathbf{Q\alpha,\mathbf{Q'\beta}}} \,,\\(\mathbf{k},\delta\rho^{\text{HF}}) &= \frac{1}{A}\sum_{\mathbf{G}}\left(\sum_{\mathbf{k'}\in\widetilde{\text{BZ}}}\sum_{\mathbf{Q''}}\sum_{\alpha'}V(\mathbf{-G})\delta\rho^{\text{HF}}_{\mathbf{(Q''+G)\alpha,\mathbf{Q''}\alpha}}(\mathbf{k'})\right)\delta_{\alpha\beta}\delta_{\mathbf{Q',Q+G}}\,, \\
    \tilde{h}^{(F)}_{\mathbf{Q\alpha,\mathbf{Q'\beta}}}(\mathbf{k},\delta\rho^{\text{HF}}) &= -\frac{1}{A}\sum_{\mathbf{G}}\left(\sum_{\mathbf{k'}\in\widetilde{\text{BZ}}}\sum_{\mathbf{Q''}}V(\mathbf{k'-k+Q''-Q})\delta\rho^{\text{HF}}_{\mathbf{(Q''+G)\beta,\mathbf{Q''}\alpha}}(\mathbf{k'})\right)\delta_{\mathbf{Q',Q+G}}\,,
\end{align}
where $A = N_{\mathbf{k},\text{HF}}A_{\text{uc}}$, with $A_{\text{uc}}$ the area of the crystal unit cell and $N_{\mathbf{k,\text{HF}}}$ is the number of $\mathbf{k}$ points used for the HF calculation.
To evaluate the non-self-consistent field at $\mathbf{k}$ points beyond the HF mesh, the summation in the parentheses runs over the $\mathbf{k}$ mesh used in the HF calculation, which we denote by $\widetilde{\text{BZ}}$ and $\delta\rho^{\text{HF}}_{\mathbf{Q}\alpha,\mathbf{Q'}\beta}$ is the converged HF solution.

\section{Hartree-Fock Results}\label{appsec:hf}

\subsection{The continuous translation symmetry breaking}
The density distribution is given by 
\begin{equation}
    \rho(\mathbf{r})=\sum_{\alpha}\braket{a^{\dagger}_{\mathbf{r},\alpha}a_{\mathbf{r},\alpha}} = \frac{1}{N_{\mathbf{k}}A_{\text{uc}}}\sum_{\mathbf{k}\in\text{BZ}}\sum_{\alpha}\sum_{\mathbf{Q,G}}\left[ \rho (\bm{k}) \right]_{\mathbf{Q\alpha},\mathbf{(Q+G)\alpha}}e^{i\mathbf{G\cdot r}} = \frac{1}{A_{\text{uc}}}\sum_{\mathbf{G}}\rho(\mathbf{G})e^{i\mathbf{G\cdot r}} \,,
\end{equation} 
which defines the order parameter $\rho(\mathbf{G})$. 
We use the average of $|\rho(\mathbf{G})|$ in the first shell to characterize the strength of continuous symmetry breaking.
Specifically, we take $|\rho({\mathbf{G}})| = 0.02$, which is approximately $10\%$ of the peak value we obtained in our parameter sweeps, as the threshold value separating the crystalline and non-crystalline phases. 
Line cut sweeps along fixed displacement field potential are shown in Fig.~\ref{fig:rho_G}(d-e).
In the insulating WC region (low displacement field), we observe a jump in the order parameter $|\rho({\mathbf{G}})|$ with increasing electron density, which signals a transition between two different insulating crystalline states. 
In the intermediate displacement field potential region, a plateau is observed over the intermediate electron density range, which corresponds to the metallic crystalline states.

We also calculate the static structure factor $S(\mathbf{q}+\mathbf{G}) = \frac{1}{N_{\mathbf{k}}A_{\mathrm{uc}}}\braket{\rho_{\mathbf{q}+\mathbf{G}}\rho_{-\mathbf{q}-\mathbf{G}}}$, where $\vq$ is the momentum component in the first Brillouin zone, $\mathbf{G}$ is the reciprocal vector and $\rho_{\mathbf{q}+\mathbf{G}} = \sum_{\mathbf{k},\alpha}a^{\dagger}_{\mathbf{k},\alpha}a_{\mathbf{k}+\mathbf{q}+\mathbf{G},\alpha}$ (note $\rho(\mathbf{q}+\mathbf{G}) = \langle \rho_{\mathbf{q}+\mathbf{G}}\rangle$). 
Within our Hartree-Fock framework, any correlation function can be expressed as products of density matrix according to Wick's theorem.
Hence, the structure factor can be written as:
\begin{align}
    S(\mathbf{q}+\mathbf{G}) =& \frac{1}{N_{\mathbf{k}}A_{\mathrm{uc}}}\sum_{\mathbf{k},\mathbf{k'}\in \text{BZ}}\sum_{\mathbf{Q},\mathbf{Q'}}\sum_{\alpha\beta}\braket{a^{\dagger}_{\mathbf{k},\mathbf{Q},\alpha}a_{\mathbf{k}+\mathbf{q},\mathbf{Q}+\mathbf{G},\alpha}a^{\dagger}_{\mathbf{k'}+\mathbf{q},\mathbf{Q'+G},\beta}a_{\mathbf{k'},\mathbf{Q'},\beta}} \nonumber\\
    =& \frac{1}{N_{\mathbf{k}}A_{\text{uc}}}\sum_{\mathbf{k},\mathbf{k'}\in\text{BZ}}\sum_{\mathbf{Q},\mathbf{Q'}}\sum_{\alpha\beta}[\braket{a^{\dagger}_{\mathbf{k},\mathbf{Q},\alpha}a_{\mathbf{k}+\mathbf{q},\mathbf{Q}+\mathbf{G},\alpha}}\braket{a^{\dagger}_{\mathbf{k'}+\mathbf{q},\mathbf{Q'}+\mathbf{G},\beta}a_{\mathbf{k'},\mathbf{Q'},\beta}}+ \braket{a^{\dagger}_{\mathbf{k},\mathbf{Q},\alpha}a_{\mathbf{k'},\mathbf{Q'},\beta}}\braket{a_{\mathbf{k}+\mathbf{q},\mathbf{Q}+\mathbf{G},\alpha}a^{\dagger}_{\mathbf{k'}+\mathbf{q},\mathbf{Q'}+\mathbf{G},\beta}}] \nonumber\\
    =& \frac{1}{N_{\mathbf{k}}A_{\mathrm{uc}}}\sum_{\mathbf{k},\mathbf{k'}\in\text{BZ}}\sum_{\mathbf{Q},\mathbf{Q'}}\sum_{\alpha\beta}
    [\rho_{\mathbf{Q\alpha},(\mathbf{Q+G})\alpha}(\mathbf{k})\rho_{(\mathbf{Q'}+\mathbf{G})\beta,\mathbf{Q'}\beta}(\mathbf{k'})\delta_{\mathbf{q},\mathbf{0}}- \rho_{\mathbf{Q}\alpha,\mathbf{Q'}\beta}(\mathbf{k})\rho_{(\mathbf{Q'}+\mathbf{G})\beta,(\mathbf{Q}+\mathbf{G})\alpha}(\mathbf{k}+\mathbf{q})\delta_{\mathbf{k},\mathbf{k'}}\nonumber\\
    &+\rho_{\mathbf{Q}\alpha,\mathbf{Q'}\beta}(\mathbf{k})\delta_{\mathbf{k,k'}}\delta_{\mathbf{Q},\mathbf{Q'}}\delta_{\alpha\beta}]  \nonumber\\
    =& \frac{N_{\mathbf{k}}}{A_{\mathrm{uc}}}\rho(\mathbf{G})\rho(-\mathbf{G})\delta_{\mathbf{q},\mathbf{0}}- \frac{1}{N_{\mathbf{k}}A_{\mathrm{uc}}}\sum_{\mathbf{k}\in\text{BZ}}\sum_{\mathbf{Q},\mathbf{Q'}}\sum_{\alpha\beta}\rho_{\mathbf{Q}\alpha,\mathbf{Q'}\beta}(\mathbf{k})\rho_{(\mathbf{Q'}+\mathbf{G})\beta, (\mathbf{Q}+\mathbf{G})\alpha}(\mathbf{k}+\mathbf{q}) + \frac{1}{N_{\mathbf{k}}A_{\mathrm{uc}}}\sum_{\mathbf{k}\in\text{BZ}}\sum_{\mathbf{Q},\alpha}\rho_{\mathbf{Q}\alpha,\mathbf{Q}\alpha}(\mathbf{k})\,.
\label{equation:structure_factor}
\end{align} 
This expression contains two types of terms: the first term is nonzero only at reciprocal lattice vectors, where it scales linearly with the system size; it therefore becomes a Dirac $\delta$ function in the thermodynamic limit. 
In contrast, the second and third terms each contain a prefactor $1/N_{\mathbf{k}}$ together with a single summation over $\mathbf{k}$ and thus remain regular in the thermodynamic limit. 
We therefore reorganize them into the following form:
\begin{align}
    & S(\vq + \mathbf{G}) = S_{\rm Bragg}(\vq + \mathbf{G}) + S_{\rm regular}(\vq + \mathbf{G}) \,,\\
    & S_{\rm Bragg}(\vq+\mathbf{G}) = \frac{N_{\mathbf{k}}}{A_{\text{uc}}}\rho(\mathbf{G})\rho(-\mathbf{G})\delta_{\mathbf{q},\mathbf{0}}\,, 
    \label{eqn:Bragg_peak}\\
    & S_{\rm regular}(\vq+\mathbf{G}) = - \frac{1}{N_{\mathbf{k}}A_{\text{uc}}}\sum_{\mathbf{k}\in\text{BZ}}\sum_{\mathbf{Q,Q'}}\sum_{\alpha\beta}\rho_{\mathbf{Q\alpha},\mathbf{Q'}\beta}(\mathbf{k})\rho_{(\mathbf{Q'+G})\beta, (\mathbf{Q+G})\alpha}(\mathbf{k}+\mathbf{q}) + \frac{1}{N_{\mathbf{k}}A_{\text{uc}}}\sum_{\mathbf{k}\in\text{BZ}}\sum_{\mathbf{Q},\alpha}\rho_{\mathbf{Q}\alpha,\mathbf{Q}\alpha}(\mathbf{k}) \,.
\end{align}
Crystalline order is characterized by the Bragg peak term $S_{\text{Bragg}}(\mathbf{G})$ at the reciprocal lattice vectors.  
Therefore, it is equivalent to use $|\rho(\mathbf{G})|$ directly characterizing the crystalline order within the HF framework.

\subsection{Band topology of the Hartree-Fock solution}
\label{appsec:hf-topology}

Using the method presented in Sec.~\ref{appsec:nonselfconsistent}, we perform the non-self-consistent calculation on a $96\times96$ momentum mesh over the first Brillouin zone and calculate the Chern number of the lowest conduction band with respect to charge neutrality by summing over the Berry flux:
\begin{equation}
    \delta\Phi(k_1,k_2) = -i\log \left[W_{(k_1+\delta k_1,k_2),(k_1,k_2)}W_{(k_1+\delta k_1,k_2+\delta k_2),(k_1+\delta k_1,k_2)}W_{(k_1,k_2+\delta k_2),(k_1+\delta k_1,k_2+\delta k_2)}W_{(k_1,k_2),(k_1,k_2+\delta k_2)}\right]\,,
\end{equation}
where $(k_1, k_2)$ stands for the two components of the wave vector defined as $\vk = \frac{k_1}{2\pi} \mathbf{b}_1 + \frac{k_2}{2\pi} \mathbf{b}_2$;
$W_{(k_1',k_2'),(k_1,k_2)}$ is the Wilson line between points $(k_1',k_2')$ and $(k_1,k_2)$, whose definition is:
\begin{equation}
     W_{(k_1',k_2'),(k_1,k_2)} = \sum_{\mathbf{Q\alpha}}u_{\mathbf{Q\alpha},n}^*(k'_1,k'_2)u_{\mathbf{Q\alpha},n}(k_1,k_2)\,. 
\end{equation}
Here $u_{\mathbf{Q\alpha},n}(k_1,k_2)$ is the single particle wave function defined in Eq.~(\ref{eqn:hf-eigen-eqn}).
The Chern number is given by the summation of the Berry flux over the momentum mesh in the first Brillouin zone 
\begin{equation}
    C = \frac{1}{2\pi}\sum_{(k_1,k_2)\in\text{BZ}}\delta\Phi(k_1,k_2)\,.
\end{equation}

\begin{figure}[t]
    \centering                    
    \includegraphics[width=0.9\linewidth]{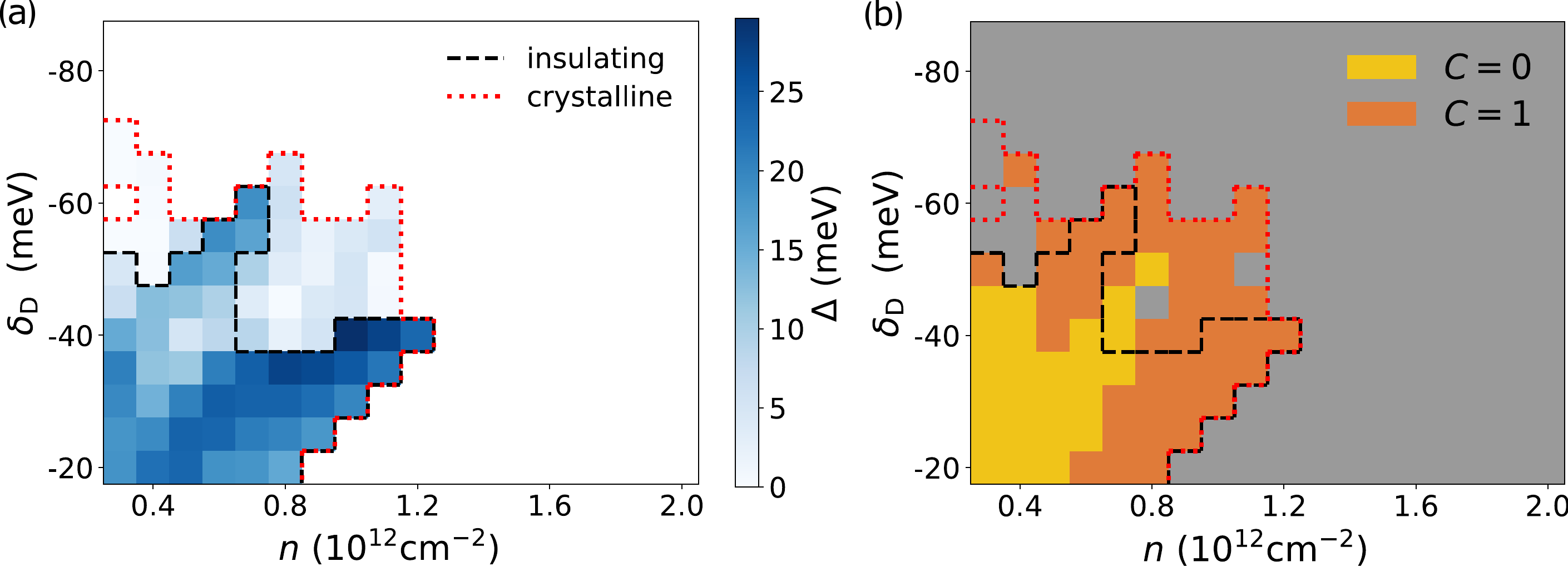} 
    \caption{(a) The direct band gap between the highest occupied and the lowest unoccupied HF bands. 
    (b) Chern number of the lowest HF band above charge neutrality. 
    In these two plots, we only present results for the crystalline states.
    We only show the Chern number when the HF solution has a direct gap $ >0.5\ \text{meV}$. 
    }
    \label{fig:M5N12_chern_number}                              
\end{figure}

To have a well defined Chern number, the band must be separated by a gap to other bands.
Fig.~\ref{fig:M5N12_chern_number}(a) shows the direct band gap between the highest occupied and the lowest unoccupied HF bands as a function of electron density $n$ and interlayer potential $\delta_{\rm D}$. 
The Chern number of the highest occupied band is evaluated when this direct gap is greater than $> 0.5 \rm meV$, and the results are shown in Fig.~\ref{fig:M5N12_chern_number}(b). 
In the low displacement field regime, we find a Chern number change from $C =0$ to $C = 1$ as we increase the density of electrons. The transition happens around $n \approx 0.6\times 10^{12}\ \text{cm}^{-2}$, which is also the position the order parameter $|\rho({\mathbf{G}})|$ jumps. 
In this transition, we do not observe a gap closing as shown in Fig.~\ref{fig:M5N12_chern_number}. 
This suggests that the $C=0$ to $C = 1$ transition is first order~\cite{Feng2026Selfdoped}.

\subsection{Free energy and the stability of 
the Hartree-Fock solutions}
\label{appsec:hf-stability}

\begin{figure}[t]
    \centering
    \includegraphics[width=1.0\linewidth]{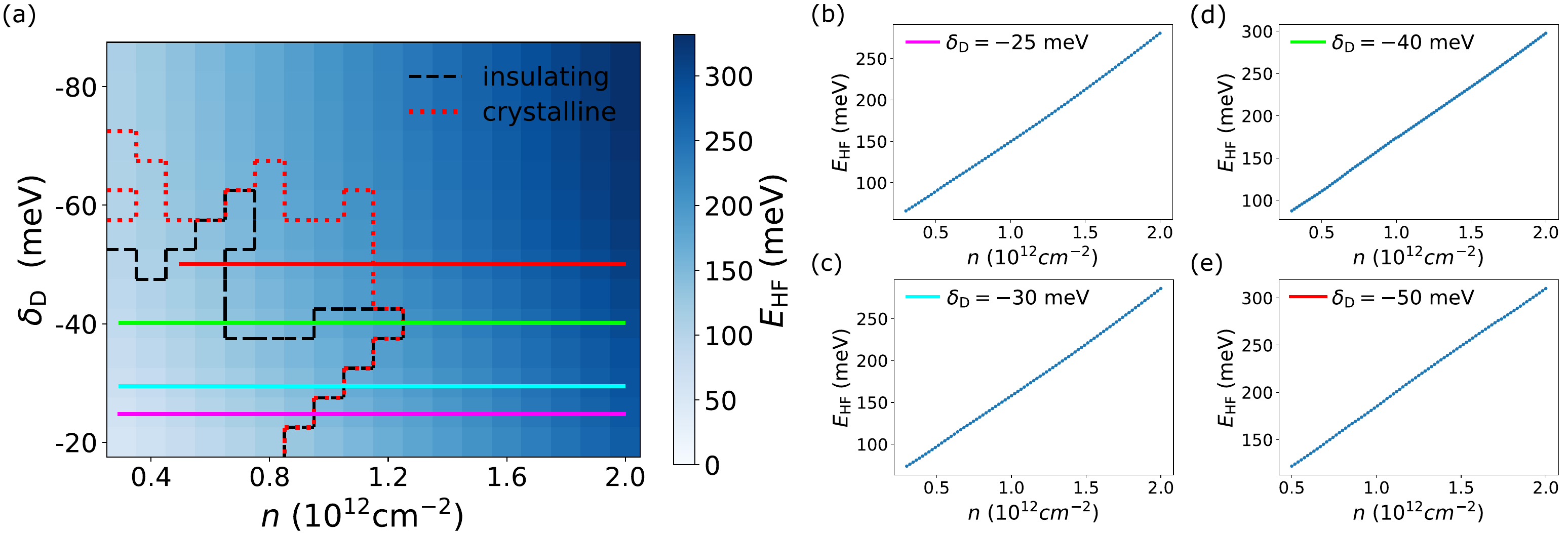}
    \caption{(a) 
    The HF energy per electron across the phase diagram. (b)-(e) The line cuts of the HF energy per electron vs. $n$ for various fixed interlayer potentials. 
    }
    \label{fig:HF_energy}
\end{figure}

\begin{figure}[t]
    \centering
    \includegraphics[width=0.9\linewidth]{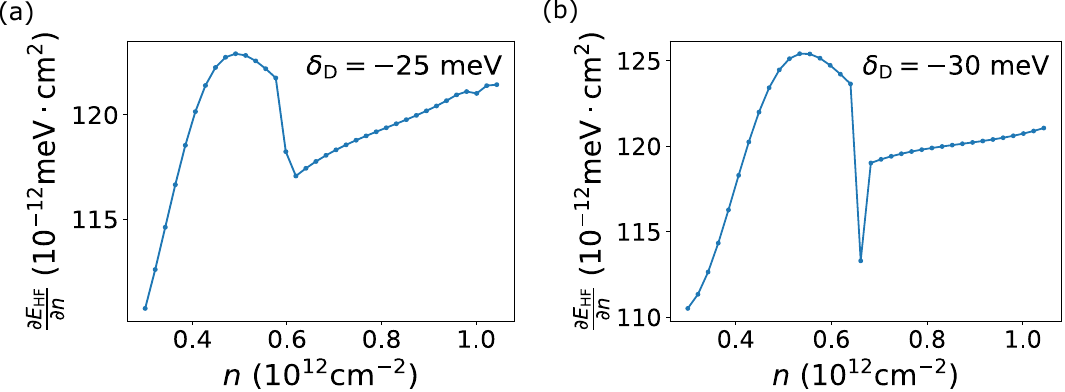}
    \caption{The first order derivative of the HF energy per electron with respect to electron densities for fixed interlayer potential (a) $\delta_{\mathrm{D}} = -25\ \text{meV}$ and (b) $\delta_{\mathrm{D}} = -30\ \text{meV}$. It shows non-analytic behavior near the $C = 0$ to $C = 1$ transition mentioned in Sec.~\ref{appsec:hf-topology}.
    }
    \label{fig:HF_energy_derivative}
\end{figure}

Here we present the free energy of the 
HF solution.
Fig.~\ref{fig:HF_energy} shows the HF energy
that varies smoothly throughout the phase diagram. In particular, along the line cuts at 
the covered interlayer potentials,
the HF energy remains continuous.
This indicates that 
none of our solutions is an artifact
of the self-consistent iteration being trapped in a local minimum.
We further observe non-analytic behavior in the first order derivative of the HF energy with respect to electron densities along various interlayer potential cuts presented in Fig.~\ref{fig:HF_energy_derivative}, consistent with the first-order phase transition mentioned at the end of Sec.~\ref{appsec:hf-topology}.

\end{document}